\documentclass[superscriptaddress,twocolumn,bibnotes,amsmath,amssymb,aps,pra,floatfix]{revtex4-2}

\usepackage{bm}
\usepackage{hyperref}
\usepackage{physics}
\usepackage{graphicx}
\usepackage{chngcntr} 
\usepackage{array}
\usepackage{multirow}

\newcommand{\be}{\begin{equation}} 
\newcommand{\ee}{\end{equation}} 
\newcommand{\nrad}{\mathcal{N}}
\begin{document}

\title{\texorpdfstring{Emergent s-wave interactions between identical fermions\\ in quasi-one-dimensional geometries}{}} 

\author{Kenneth G.\ Jackson}
\affiliation{Department of Physics and CQIQC, University of Toronto, Toronto, Ontario M5S 1A7, Canada}
\author{Colin J.\ Dale}
\affiliation{Department of Physics and CQIQC, University of Toronto, Toronto, Ontario M5S 1A7, Canada}
\author{Jeff Maki}
\affiliation{Department of Physics and HKU-UCAS Joint Institute for Theoretical and Computational Physics at Hong Kong, The University of Hong Kong, Hong Kong, China}
\author{Kevin G.\ S.\ Xie}
\affiliation{Department of Physics and CQIQC, University of Toronto, Toronto, Ontario M5S 1A7, Canada}
\author{Ben A.\ Olsen}
\affiliation{Yale-NUS College, Singapore 138527, Singapore}
\author{Denise J.\ M.\ Ahmed-Braun}
\affiliation{Department of Physics, Eindhoven University of Technology, The Netherlands}
\author{Shizhong Zhang}
\affiliation{Department of Physics and HKU-UCAS Joint Institute for Theoretical and Computational Physics at Hong Kong, The University of Hong Kong, Hong Kong, China}
\author{Joseph H.\ Thywissen}
\affiliation{Department of Physics and CQIQC, University of Toronto, Toronto, Ontario M5S 1A7, Canada}

\date{\today}

\begin{abstract}
Orbital degrees of freedom play an essential role in metals, semiconductors, and strongly confined electronic systems. Experiments with ultracold atoms have used highly anisotropic confinement to explore low-dimensional physics, but typically eliminate orbital degrees of freedom by preparing motional ground states in strongly confined directions. 
Here we prepare multi-band systems of spin-polarized fermionic potassium ($^{40}$K) in the quasi-one-dimensional (q1D) regime and quantify the strength of atom-atom correlations using radio-frequency spectroscopy. 
The activation of orbital degrees of freedom leads to a new phenomenon: a low-energy scattering channel that has even particle-exchange parity along the q1D axis, as if the underlying interactions were s-wave. This emergent exchange symmetry is enabled by orbital singlet wave functions in the strongly confined directions, which also confer high-momentum components to low-energy q1D collisions. 
We measure both the q1D odd-wave and even-wave ``contact'' parameters for the first time, and compare them to theoretical predictions of one-dimensional many-body models. 
The strength and spatial symmetry of interactions are tuned by a p-wave Feshbach resonance and by transverse confinement strength. Near resonance, the even-wave contact approaches its theoretical unitary value, whereas the maximum observed odd-wave contact remains several orders of magnitude below its unitary limit. 
Low-energy scattering channels of multi-orbital systems, such as those found here, may provide new routes for the exploration of universal many-body phenomena. 
\end{abstract}

\maketitle

\section{Introduction \label{sec:intro}}

Strong confinement of many-body systems has proven to be a fruitful approach for the creation of novel and emergent quantum states. For instance, the confinement of an electron gas in a two-dimensional quantum well with a strong perpendicular magnetic field leads to the manifestation of the integer and fractional quantum Hall effects, which exhibit exceptional transport properties~\cite{prange2012qhe}. Similarly, relaxing the electron spin degrees of freedom can result in the emergence of novel excitations such as skyrmions~\cite{Barrett1995,Aifer1996}. In ultracold atomic gases, recent advances in the manipulation of quantum gases in low dimensions have led to similar progress. For instance, fractional spin and charge degrees of freedom have been observed in atomic gases under strong transverse confinement~\cite{Gross:2020,Hulet:2022}. Moreover, confinement-induced resonances~\cite{Olshanii:1998} can be used to change the interaction between atoms, providing a platform to investigate many-body physics in a disorder-free environment \cite{Giamarchi:2017el}. 

In studies with cold atoms, research has typically focused on regimes where transverse excitations along the strong confinement directions are irrelevant, with some exceptions such as Ref.~\cite{Schmiedmayer:2021}. For instance, in a highly elongated trap, it is commonly assumed that the atoms occupy the lowest transverse oscillator state, such that the only active degrees of freedom are those along the weakly confined (axial) direction. 
As a result, three-dimensional (3D) systems of spin-polarized bosons and spin-half fermions under such confinement have been successfully modeled as quasi-one-dimensional (q1D) systems whose interactions are {\em even-wave}, i.e., described by one-dimensional (1D) pair wave functions that have exchange-even symmetry  \cite{Olshanii:1998,Bergeman:2003kn,Moore:2004,Schmelcher:2005,Yurovsky:2005,Naidon:2007,Dunjko:2011,Kristensen:2015ib}.  

Here we investigate the role of transverse orbital degrees of freedom in a simplified setting: a spin-polarized Fermi gas of $^{40}$K with p-wave interactions confined in a q1D geometry. By eliminating the spin degree of freedom, the possibility of spin-singlet s-wave scattering is excluded, so that the role of orbital degrees of freedom is clearly exhibited. Our work builds on prior studies of {\em odd-wave} interactions, i.e., described by one-dimensional (1D) pair wave functions that have exchange-odd symmetry. Experimental work with low-dimensional p-wave systems have suppressed collisions along the strongly confined directions \cite{Gunter:2005er,Waseem:2016ki,Chang:2020,Marcum:2020}. Q1D p-wave systems have been the subject of numerous theoretical studies  \cite{Granger:2004im,Kanjilal:2004jv,Girardeau:2004,Pricoupenko:2008fq,Hess:2014hd,Gao:2015hz,Saeidian:2015db,Zhou:2017jv,Sekino:2018gk,Maki:2021,Tanaka2022,Nishida2022,Jeff2022} including a proposal to enable a p-wave halo dimer state~\cite{Fonta:2020ev} and to use strong confinement to reduce the three-body recombination loss~\cite{Levinsen:2008,Kurlov:2017} that plagues 3D p-wave gases \cite{Regal:2003go,Zhang:2004cy,Schunck:2005cf,Chevy:2005im,Inada:2008hz}. 

\begin{figure*}[t!]
\centering
\includegraphics[width=1.75\columnwidth]{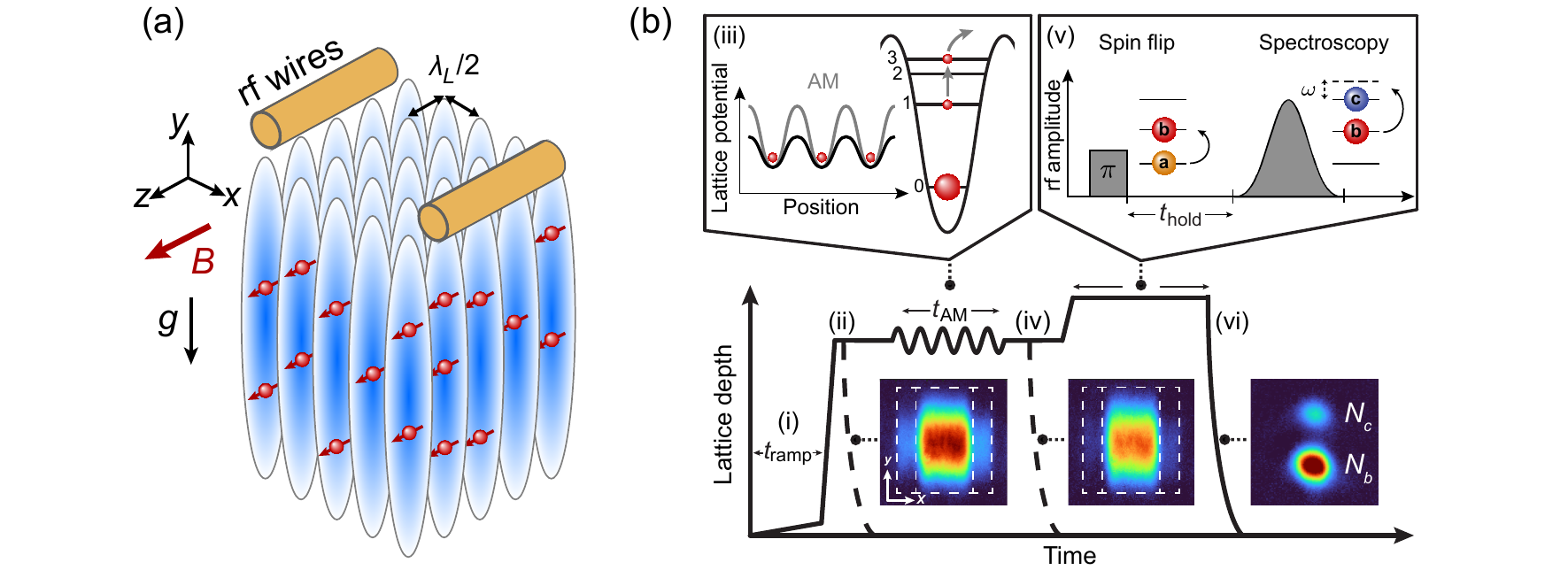}
\caption{{\em Experimental overview.} 
(a) An array of q1D traps aligned in the vertical ($y$) direction is created by two orthogonal standing waves of light at wavelength $\lambda_L$. A static magnetic field $B$ is oriented along $z$. Radio-frequency fields are generated by a pair of proximate microfabricated conductors. 
(b) Experimental sequence, depicting the potential depth of the confining two-dimensional optical lattice versus time. 
(i) Following the creation of a degenerate Fermi gas, the optical standing waves are ramped on to generate the array of q1D traps. The two-part ramp includes a slow increase (in $t_\mathrm{ramp}$) to minimize heating followed by a fast jump to to isolate q1D ensembles.
(ii) At this stage, time-of-flight band mapping typically indicates that the majority of atoms are in the ground band (central dashed box) but some atoms are also loaded into the first excited band (lateral boxes).
(iii) The lattice depth is optionally modulated for $t_\mathrm{AM}$ before being ramped to its full depth. The frequency is chosen to drive excited-band atoms resonantly into the third excited band, from which they leave the trap.
(iv) Time-of-flight images confirm that AM reduces population in the excited band. 
(v) Feshbach-enhanced p-wave interactions are activated by a spin flip from $\ket{a}$ to $\ket{b}$. After a time $t_\mathrm{hold}$ the system is probed with a spectroscopic pulse at detuning $\omega$. 
(vi) The fraction of atoms transferred from $\ket{b}$ to $\ket{c}$ by the spectroscopic rf pulse is measured with state-selective imaging. 
\label{fig:methods} }
\end{figure*}

In our experiment, we activate the orbital degree of freedom through controlled population of the first excited transverse band. We discover conditions under which fermions interact through wave functions with even particle-exchange symmetry for the q1D degree of freedom along the axial direction. We explain how fermionic symmetry is preserved by exchange-antisymmetric transverse orbital states. 
The magnetic Feshbach resonance used to tune p-wave interactions also adds an anisotropic nature to scattering. 
This vector character determines the magnetic field values that can host emergent s-wave resonances and can be used to unambiguously identify the nature of the observed interactions. 

We find clear signatures of interaction-induced correlations between atoms through rf spectroscopy for both odd- and even-wave scattering along the axial direction. Experimental investigations are supplemented by a theoretical analysis that re-examines the two-body problem and presents a unified treatment for arbitrary energies when multiple transverse bands are relevant. This allows us to describe emergent even-wave scattering on equal footing with odd-wave scattering and to relate this to novel correlations observed in rf spectroscopy. Specifically, we relate the high-frequency limit of the rf transfer rate between two internal spin states to the short-range correlations between atoms through universal contact relations~\cite{Tan:2008a,Tan:2008b,Tan:2008c,Braaten:2008tc,Braaten:2008ez,Werner:2009vb,Zhang:2009kq,Braaten2010,Braaten:2012gh,Werner:2012um,Werner:2012hy,Yoshida:2015hh,Yu:2015go,Luciuk:2016gr,He:2016bn,HuiHu:2016}. 
Our theoretical treatment delineates the conditions under which the rf spin-flip rate is a measure of the 1D contact parameter. Notably, the rf spectra match true 1D scaling up to $2\omega_\perp$, where $\hbar \omega_\perp$ is the single-particle band gap.  
We extract the odd-wave contact parameter anticipated by Refs.~\cite{He:2016gh,Cui:2016kd,Cui:2016ho,Sekino:2018gk,Yin:2018hl,Maki:2021} for the first time, and also measure the even-wave contact parameter. 
A distinguishing feature of the latter is a novel singular structure at $2\omega_\perp$ that is anticipated by our analysis. 

Our work suggests that introducing energy-gapped orbital degrees of freedom can be used more broadly to engineer the low-energy exchange symmetry of few-body and many-body systems. Our theoretical analysis provides a framework for understanding how these novel orbit-dressed collision channels emerge from the one- to three-dimensional crossover. Both the conceptual paradigm and the measurement protocols introduced here could be expanded to bosonic systems, spin mixtures, quasi-two-dimensional systems, and alternative partial waves. 

The manuscript is structured as follows. 
In Sec.~\ref{sec:exp}, we provide an overview of the experimental protocol. 
In Sec.~\ref{sec:q1Dstates}, we describe p-wave scattering in the presence of strong q1D confinement and orbital degrees of freedom. Adapted to the experimental scenario, our treatment identifies three scattering continua with distinct pair wave functions. 
In Sec.~\ref{sec:oddExp}, we measure the q1D odd-wave correlation strength and delineate the range of confinement strengths and magnetic fields in which the contact parameter is governed by the q1D odd-wave scattering length. We also demonstrate that the maximum contact parameter remains two orders of magnitude below its expected unitary value. 
In Sec.~\ref{sec:evenExp}, we present three experimental investigations of emergent s-wave interactions. The even-wave contact parameter is seen to approach its expected unitary value, and we estimate that the inferred interaction strength is at least an order of magnitude stronger than that of odd-wave interactions. 
We conclude in Sec.~\ref{sec:conclusion} with a summary of key findings and a discussion of future prospects enabled by multi-orbital scattering in low-dimensional systems. 

\section{Experimental protocol and summary of observations \label{sec:exp}}

Our ensemble of q1D systems is prepared as follows. A bulk sample of $^{40}$K, spin-polarized in the lowest hyperfine-Zeeman state $\ket{a}$, is confined in an optical trap and cooled sympathetically with bosonic $^{87}$Rb. After the final cooling stage, $N_\mathrm{tot} = 7(1)\times 10^4$ spin-polarized $^{40}$K atoms at $T \sim 300$\,nK are loaded into a lattice of tube-like anisotropic traps [see Fig.~\ref{fig:methods}(a)]. The confinement along the $x$ and $z$ directions is created by two orthogonal standing waves, produced with retro-reflected beams of light with wavelength $\lambda_L=760.6$\,nm. 

In the deep-lattice limit where tunneling between adjacent tubes can be neglected, each lattice site is harmonic near its minimum, with longitudinal oscillation frequency $\omega_y \approx 2 \pi \times 400$\,Hz. The transverse confinement in turn is set by the lattice depth $V_L$, measured in units of the recoil energy, $E_R = h^2/(2 m \lambda_L^2) \approx h \times 8.6$\,kHz, where $m$ is the atomic mass of $^{40}$K. With confinement in both transverse directions set to $V_L=40\,E_R$, for instance, the transverse oscillation frequency is $\omega_\perp \approx 2 \pi \times 100$\,kHz. 
The transformation of the 3D cloud into an array of q1D clouds begins with a ramp of the lattice intensity, across a duration $t_\mathrm{ramp}$ (typically $300$\,ms), to a modest depth, less than $10 E_R$ [Fig.~\ref{fig:methods}(b)(i)]. Tubes are then isolated with a more rapid ramp to $40(2)\,E_R$. We use the loading model described in App.~\ref{app:inhomogeneity} to estimate the final number of atoms and temperature distribution in the tubes. For typical experimental parameters, the peak number per tube is $N \approx 30$ and $T/T_F \approx 0.8$, where $T_F$ is the 1D Fermi temperature. 

Band populations are measured using band-mapped time-of-flight imaging. We adjust the chemical potential of the gas by tuning the optical trap depth and thereby optimize for either ground-band loading or multi-band loading. For the latter, when the chemical potential of the bulk gas is comparable to $E_R$, some atoms are loaded into the first excited state of each transverse direction since loading at the band edge cannot be adiabatic [Fig.~\ref{fig:methods}(b)(ii)]. The fraction of atoms loaded into the first excited state is typically $15\%$ in each direction. 

To prepare for dedicated experiments measuring odd-wave correlations without orbital excitations (Sec.~\ref{sec:oddExp}), we deplete the excited-band population using amplitude modulation (AM). The power of both optical standing waves is modulated for a time $t_\mathrm{AM}$ at a frequency resonant with the transition from the single-atom first-to-third excited bands [see Fig.~\ref{fig:methods}(b)(iii)]. Due to the anharmonicity of the standing-wave confinement, this modulation is off-resonant from the zero-to-two excitation, leaving the number of atoms in the lowest band relatively unaffected. When optimizing for ground-band loading, band mapping after AM [Fig.~\ref{fig:methods}(b)(iv)] indicates that the excited-band population is reduced to $\lesssim 5\%$ in each lattice direction. Lattice depth calibration is performed via AM of the single-particle zero-to-two excitation, with $5\%$ uncertainty. Ensembles for even-wave correlation measurements (Sec.~\ref{sec:evenExp}) are prepared without AM, or with phase modulation of the lattice to enhance population in the first excited bands (Sec.~\ref{app:bandpop}). 

With either preparation, the q1D ensembles of atoms in the lowest hyperfine-Zeeman state $\ket{a}$ are long-lived and weakly interacting. A p-wave Feshbach resonance \cite{Regal:2003go,Zhang:2004cy,Schunck:2005cf} exists in the $bb$ collision channel, where $\ket{b}$ is the second-lowest hyperfine Zeeman state, adiabatically connected to the $\ket{F=9/2, m_F = -7/2}$ state at low magnetic field. After setting the field $B$ and lattice depth $V_L$ for a particular measurement, interactions are snapped on by transferring the atoms to the state $\ket{b}$ using a $40$-$\mu$s-long rf pulse [see Fig.~\ref{fig:methods}(b)(v)]. 

After a hold time $t_\mathrm{hold}$ of typically $160\,\mu$s, a spectroscopic pulse of duration $\tau = 120\,\mu$s is applied. The frequency is offset $\omega$ above the single-particle $\ket{b}$-to-$\ket{c}$ resonant frequency, where $\ket{c}$ is the third-lowest hyperfine-Zeeman state (and assumed to be weakly interacting: see App.~\ref{app:finalstate}). The number of atoms transferred, $N_c$, is assessed by state-sensitive imaging in time-of-flight, shown in Fig.~\ref{fig:methods}(b)(vi).  The spectroscopic pulse has a Blackman envelope to minimize Fourier broadening, such that the measured transfer rate $\Gamma$ is given by $N_c/\tau$ times a corrective envelope factor. 

\begin{figure*}[t]
\centering
\includegraphics[width=2\columnwidth]{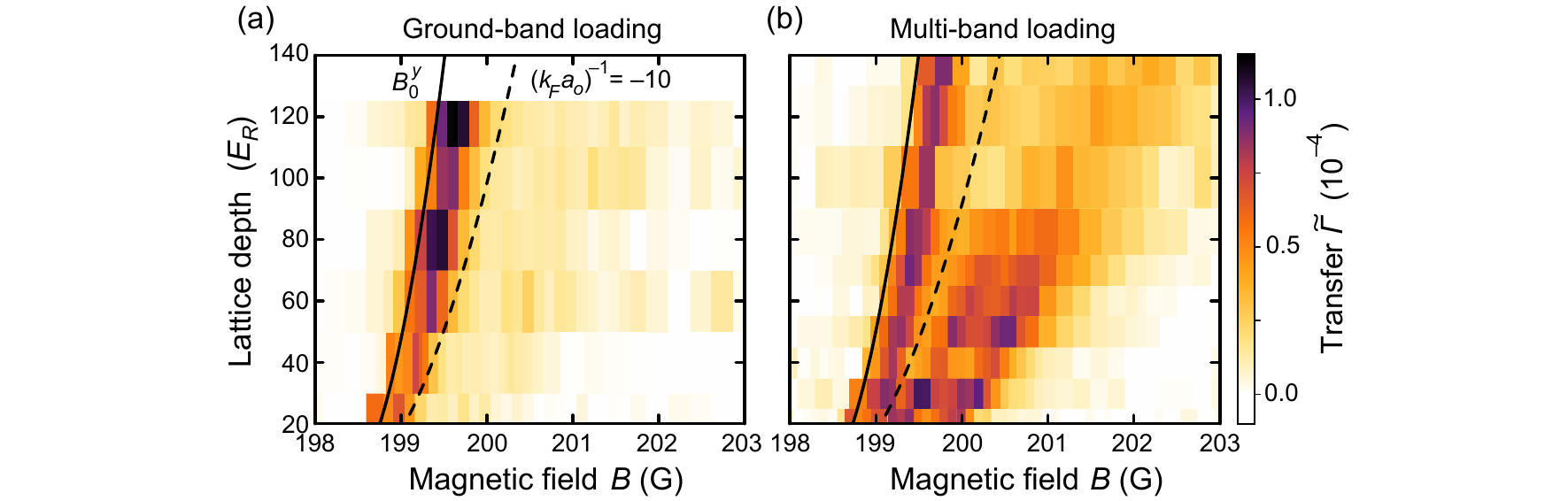}
\caption{{\em Overview of observed correlations.} 
The normalized rf transfer rate $\widetilde\Gamma$, which is a measure of atom-atom correlation strength, is indicated by color across a range of magnetic fields and optical lattice depths. 
The depth of the confining optical lattice, $V_L$, is given in units of the recoil energy $E_R = h^2/2m\lambda_L^2$. The first peak in $\widetilde\Gamma$ lies between the odd-wave resonance (indicated by a solid black line at $B_{0}^{y}$, where $a_o^{-1} = 0$ as discussed in Sec.~\ref{sec:q1Dstates}) 
and $(k_F a_o)^{-1} = -10$ (indicated by a black dashed line).
(a) Without active orbital degrees
of freedom, atom-atom correlations are found only within this range. (b) With multi-band loading, correlations beyond the odd-wave regime are found. We show that these additional correlations are due to even-wave interactions along the q1D direction. The structure of $\widetilde{\Gamma}(B,V_L)$ is further explored in Figs.~\ref{fig:oddcontact}, \ref{fig:vector}, and \ref{fig:evencontact}.
\label{fig:results} }
\end{figure*}

Figure~\ref{fig:results} provides a side-by-side comparison of correlations found with and without activated orbital degrees of freedom. 
The normalized rf transfer rate $\widetilde\Gamma = E_F \Gamma /( \hbar \pi \Omega^2 N_\mathrm{tot})$ is shown versus lattice depth and magnetic field, where $\Omega$ is the Rabi frequency and $E_F$ is the calculated Fermi energy of the central tube.
As will be discussed in Section~\ref{ssec:Tmatrix}, $\Gamma$ is proportional to short-range correlations, which are quantified by so-called contact parameters \cite{Tan:2008a,Tan:2008b,Tan:2008c,Braaten:2008tc,Braaten:2008ez,Werner:2009vb,Zhang:2009kq,Braaten2010,Braaten:2012gh,Werner:2012um,Werner:2012hy,Yoshida:2015hh,Yu:2015go,Luciuk:2016gr,He:2016bn,HuiHu:2016} 
and have been the object of prior experimental studies in 3D for both s-wave and p-wave systems \cite{Randeria2010,stewart:2010,kuhnle:2010,Barth:2011,hoinka:2013,Luciuk:2016gr}
and in quasi-two-dimensional systems with s-wave interactions \cite{Frohlich:2012,Luciuk:2017}.

Without active orbital degrees of freedom, atom-atom correlations 
are found just above the odd-wave resonance ($B_0^y$ as defined later) indicated by a solid line in Figure~\ref{fig:results}(a). At $B_0^y$, the q1D odd-wave scattering length $a_o$ is predicted to diverge, due to interactions along the longitudinal direction.
The observed normalized transfer rate, $\widetilde\Gamma$, is largest at magnetic fields between $B_0^y$ and where $a_o = -(10 k_F)^{-1}$ (dashed line), where $k_F$ is the Fermi momentum in the central tube. 
In Sec.~\ref{sec:oddExp}, we show that the observed correlations agree well with those expected from weakly attractive odd-wave interactions. 

Figure~\ref{fig:results}(b) displays the same quantities, but with orbital degrees of freedom activated. 
The normalized rf transfer rate, $\widetilde\Gamma$, now reveals that atom-atom correlations appear well beyond the odd-wave parameter range (still bracketed by the solid and dashed lines).
In Sec.~\ref{sec:evenExp}, we provide clear assignment of these correlations to even-wave interactions and compare their strength to that predicted by theory. 

\section{p-wave scattering in a quasi-one-dimensional trap \label{sec:q1Dstates}}

Low-energy scattering in the q1D trap has been considered previously for even-wave scattering \cite{Olshanii:1998,Bergeman:2003kn,Moore:2004,Schmelcher:2005,Yurovsky:2005,Naidon:2007,Dunjko:2011,Kristensen:2015ib} of both identical bosons and spin mixtures of fermions, as well as for for odd-wave scattering \cite{Granger:2004im,Kanjilal:2004jv,Girardeau:2004,Pricoupenko:2008fq,Hess:2014hd,Gao:2015hz,Saeidian:2015db,Zhou:2017jv} of identical fermions. These treatments consider atoms to reside in the ground state of the transverse confinement, such that higher-energy orbital states are only relevant as closed channels \cite{Gao:2015hz}. 
Our experimental scenario requires a generalization that includes population of asymptotic states with transverse orbital excitation. When these additional orbital states are hybridized with 3D p-wave dimer states, whose inter-nuclear separation is typically much smaller than the transverse oscillator length, the new scattering resonances which emerge include both even- and odd-wave q1D symmetries. 

\begin{table}[b!]
  \caption{Summary of notation used to parameterize scattering in this work. 
  \label{tab:params}}
  \begin{ruledtabular}
  \begin{tabular}{@{\hskip 0.5cm}ccc@{\hskip 0.5cm}}
    \multirow{2}{*}{\centering resonance type} & scattering & scattering\\
     & amplitude & parameters
    \\ \hline
   3D p-wave & $f_\textrm{3D}$& $V_\textrm{3D}$, $R_\textrm{3D}$ \\
    1D odd-wave & $f_o$& $a_o$, $r_o$ \\ 
    1D even-wave & $f_e$ & $a_e$, $r_e$ \\
  \end{tabular}
  \end{ruledtabular}
\end{table}

To obtain the effective scattering parameters in q1D, let us first recall that the p-wave scattering amplitude in 3D, $f_\mathrm{3D}$, is given by
\be \label{eq:f3D}
-(k f_\mathrm{3D})^{-1} = i + (V_\mathrm{3D} k^3)^{-1} + (k R_\mathrm{3D})^{-1} + \mathcal{O}(k) \,, \ee
where $V_\mathrm{3D}$ and $R_\mathrm{3D} > 0$ are the p-wave scattering volume and effective range, respectively, and $\hbar k$ is the relative momentum. Near a p-wave Feshbach resonance, $V_{\mathrm{3D}}$ is widely tunable~\cite{Gaebler:2007,Fuchs:2008ka,AhmedBraun:2021}, but the Wigner bound \cite{Hammer:2009bd,Hammer:2010hm,Braaten:2012go} constrains the maximum value of $R_\mathrm{3D}$ to be on the order of the van der Waals length, which is small compared to typical $k^{-1}$ in the ultracold regime. Thus $(k R_\mathrm{3D})^{-1}\gg 1$ and, according to Eq.~\eqref{eq:f3D}, remains numerically important for 3D p-wave resonant scattering. In fact, it is crucial to include the effective-range term in Eq.~\eqref{eq:f3D} to describe many physical properties of p-wave Fermi gases appropriately  \cite{Bertulani02,Bedaque:2003,Gubbels:2007ig,Braaten:2012go,Luciuk:2016gr,Yu:2015go,Maki2020,Ding2019}. 

Now, let us consider a Fermi gas of identical atoms, such as $^{40}$K in a single hyperfine-Zeeman state, confined in a highly anisotropic potential. The single-particle Hamiltonian within a single tube is, neglecting anharmonic corrections, 
\begin{equation} \label{eq:H1}
H_0=\frac{1}{2m}{\bm{p}}^2 +\frac{1}{2}m(\omega_x^2 x^2+\omega_y^2 y^2+\omega_z^2 z^2)\,,
\end{equation}
where ${\bm p}=(p_x,p_y,p_z)$ is the 3D momentum of the atom. Quasi-one-dimensional confinement is achieved when the transverse harmonic frequencies in the $x$ and $z$ direction are much larger than in the $y$ direction: $\{\omega_x,\omega_z\}\gg \omega_y$. This allows a separation of energy scales, in which the collision energy (determined by the 1D Fermi energy $E_F$ at low temperature, $T \lesssim E_F/k_B$) is much smaller than the energy quanta of the transverse motion, namely $E_F \ll \{ \hbar\omega_x, \hbar\omega_z \}$. For a true 1D system of $N\gg 1$ spinless fermions, the Fermi energy is $E_F = N\hbar\omega_y$, so that the condition for q1D collisions is 
\be \label{eq:q1D}
N\lambda\ll 1\,, 
\ee
where $\lambda \equiv \omega_y/\omega_\perp$ is the trap aspect ratio, written in the case of cylindrically symmetric confinement, $\omega_x=\omega_z\equiv \omega_\perp$. For our typical confinement strength and atom numbers (see Sec.~\ref{sec:exp}), $N \lambda \sim 0.1$, i.e., well within the q1D regime. 

\subsection{Scattering in q1D regime \label{ssec:Tmatrix}} 

For non-interacting atoms of equal mass, the two-body Schr\"{o}dinger equation can be separated into center-of-mass (CM) and relative (rel) motion, each described by a Hamiltonian with the same form as Eq.~\eqref{eq:H1} except for the replacement of the mass $m$ with the total mass $2m$ or the reduced mass $\mu=m/2$, respectively \cite{Haller2010,Sala:2012,Sala:2013}%
\footnote{In q1D s-wave systems, the anharmonicity of the transverse confinement potential was observed \cite{Haller2010,Sala:2013} to induce couplings between the center-of-mass motion and the relative motion. This created new scattering resonances to bound states with CM excitation in the strongly confined direction \cite{Sala:2012}. However, we did not find any evidence of this effect in our data}%
. In the limit $N \gg 1$, we can treat the relative momentum $k$ along the q1D axis ($y$) as continuous, but retain the quantization of the transverse motion. Note that $\hbar$ has been set to unity here and for the remainder of this subsection.

In the CM frame, setting $\omega_{x}=\omega_{z} = \omega_\perp$ and $\omega_y=0$, we can write the two-body Hamiltonian as 
\begin{equation}
H_\mathrm{0,2b} = \frac{1}{2 \mu}{\bm{ p}_\textrm{rel} ^2}+ \frac{1}{2} \mu \omega_{\perp}^2 \rho^2\,,
\end{equation}
where $\bm{ p}_\textrm{rel}$ is the relative momentum and $\rho=\sqrt{x^2+z^2}$ (and $\phi$ below) are the standard cylindrical coordinates. The non-interacting relative wave function can be characterized conveniently by the principal quantum number in the radial direction, $\nrad$, and the azimuthal quantum number, $m$. Up to a normalization factor, the transverse spatial wave functions of the non-interacting eigenstates are 
\begin{equation} 
\Psi_{\nrad,m}(\rho,\phi) =  \frac{ e^{i m \phi} \rho^{|m|}} {a_{\perp}^{|m|+1}} e^{-\frac{\rho^2}{2 a_{\perp}^2}} \mathcal{L}^{|m|}_\nrad \! ({\rho^2}/{a_{\perp}^2})\,,
\label{eq:basis}
\end{equation}
where $\mathcal{L}_\nrad^{|m|}(x)$ is the generalized Laguerre polynomial and $a_{\perp}= \sqrt{1/\mu \omega_\perp}$ is the radial oscillator length. 
Including the kinetic energy in the 1D direction, $k^2/(2\mu)$, the energies of the full $\ket{\nrad,m,k}$ states are
$E_{\nrad,m,k} = \frac{k^2}{2\mu} + (2\nrad + |m|+1) \omega_{\perp}$, 
which we will also write in dimensionless form as 
\be \mathcal{E}_m \equiv \frac{E_\mathrm{rel}-(|m|+1)\omega_{\perp}}{2\omega_{\perp}} = \left(\frac{ka_{\perp}}{2}\right)^2 \,,\ee
where $E_\mathrm{rel}$ is the total energy that does not include the CM motion or CM confinement. 

Now, let us include the inter-atomic interaction described in detail in App.~\ref{app:theory}. We assume for now that the interaction is cylindrically symmetric, such that $m$ remains a good quantum number; the extension to anisotropic interactions is discussed in Sec.~\ref{ssec:anisotropy}. For an underlying p-wave interaction, interacting eigenstates can be sorted into $m=-1$, $m=0$, and $m=+1$, and will in general be a superposition of $\ket{\nrad,m,k}$ states with different $\nrad$. 

The symmetry of the asymptotic wave function along the axial direction can be seen in the emergent 1D wave functions, which are solely a function of $y$. For an incident wave with momentum $k$ in continuum $\nrad$ and with symmetry $m$, the scattered wave component with quantum numbers $\nrad'$ and $m'$ is given by 
\begin{align}
&\psi_{\nrad',m',k}(y) = \delta_{m',m}\left[\delta_{\nrad',\nrad} \, e^{i k y} \phantom{\frac{1}{1}^1}\right.\nonumber \\
&\left.\qquad + f_{\nrad'}^{(m)}  e^{-\frac{2}{a_{\perp}} \sqrt{\nrad'-\mathcal{E}_m-i\varepsilon}\,|y|} \left(\frac{y}{|y|}\right)^{\! |m|+1} \right]\,,
\label{eq:wf}
\end{align}
where $\varepsilon$ is an infinitesimal positive quantity. Here $f_{\nrad'}^{(m)}$ are the scattering amplitudes into the $\nrad'$ radial state, 
\begin{align} \label{eq:fn}
&f_{\nrad'}^{(0)} = -i \mu k \, T_o \\
\mbox{and}\quad &f_{\nrad'}^{(\pm1)} = -i \mu \frac{a_{\perp}}{2} \frac{\sqrt{\nrad'+1}\sqrt{\nrad+1}}{\sqrt{\nrad'-\mathcal{E}_1-i\varepsilon}} \, T_e\,, \nonumber
\end{align}
parameterized by odd-wave and even-wave scattering transition matrices ($T$-matrices) $T_o$ and $T_e$, which are defined below.

For the $m=0$ sector, the two-body wave function is exchange-symmetric along the transverse direction, as is evidenced by Eq.~\eqref{eq:basis} under parity transformation $\phi\to \phi+\pi$. 
The overall exchange antisymmetry of the two-body wave function then requires the axial wave function to be {\em antisymmetric} under exchange, as is observed in the scattered wave of Eq.~\eqref{eq:wf}. 
Thus two particles undergo odd-wave scattering along the q1D direction (also called 1D p-wave scattering). 

Consider the incident state $\ket{\nrad=0,m=0,k}$ just above the lowest non-interacting band energy,  such that $\mathcal{E}_0\ll 1$. The interacting wave function consists of all $\ket{\nrad,m=0,k}$ states, starting with $\nrad=0$, for which both atoms are in the single-particle ground-state orbital of the transverse confinement potential. From Eq.~\eqref{eq:wf}, we see that for higher-energy states with $\nrad' > 0$, the associated evanescent scattering wave vanishes for $|y|\gg a_\perp$. Therefore, only the component with $\nrad'=0$ and $m=0$ survives asymptotically, a state that describes two particles in the single-particle ground state of the transverse oscillator. 

For the $m=1$ and $m=-1$ sectors, the transverse wave function is odd [see again Eq.~\eqref{eq:basis} for $\phi\to \phi+\pi$] such that the {\em scattering wave function is exchange-symmetric} along $y$, the longitudinal direction 
\cite{Nishida:2010}\footnote{A complementary phenomenon can occur for scattering in mixed dimensions~\cite{Nishida:2010}. In that case, the odd parity of the CM motion can induce p-wave scattering in the relative coordinate, even though the underlying scattering is s-wave, because two atoms see different trapping potentials. In our case, the CM motion is irrelevant; instead, the odd parity of the relative motion in the strongly confined direction allows for the emergence of the even-wave scattering of identical fermions along the q1D direction.}%
.
Consider an incoming state $\ket{\nrad=0,m=1,k}$, with collision energies just above the lowest-energy $m=1$ band, such that $\mathcal{E}_1\ll 1$. Similarly to the $m=0$ sector, we see from Eq.~\eqref{eq:wf} that all the $\nrad'> 0$ components of the scattered wave function decay at $|y|\gg a_\perp$. Therefore, the asymptotic q1D wave function features a transverse component that is characterized by $\nrad=0$ and $m=1$. Similar arguments hold for the $m=-1$ sector.

The separation between the odd- and even-wave channels in the presence of interactions leads to two low-energy scattering matrices.
For $m=0$, and in the low-energy limit ($0<\mathcal{E}_0= (E_{\rm rel}-\omega_\perp)/2\omega_\perp \ll 1$),  
we find that the effective 1D scattering $T$-matrix $T_o$ is given by 
\begin{align}
T^{-1}_o &= \frac{2\mu }{2} \left[\frac{1}{a_o} + k^2 r_o +i k \right]\,, \quad \mbox{with} \nonumber \\
\frac{1}{a_o} &= \frac{a_{\perp}^2}{6}\left( \frac{1}{V_\mathrm{3D}} + \frac{2}{R_\mathrm{3D} a_{\perp}^2}\right) - \frac{2}{a_{\perp}} \zeta(-1/2) \nonumber \\
\mbox{and}\quad r_o &= \frac{a_{\perp}^2}{6R_\mathrm{3D}} + \frac{a_{\perp}}{4} \zeta(1/2)\,,
\label{eq:low_odd}
\end{align}
where $\zeta(s)\equiv \zeta(s,1)$ is the Riemann zeta function, $a_o$ is the 1D odd-wave scattering length, and $r_o > 0$ is the 1D odd-wave effective range. The key steps in our derivation are described in App.~\ref{app:theory}. These equations define the effective 1D parameters for the odd-wave scattering amplitude $f_{o}$ as
\be \label{eq:fo1D}
-f_{o}^{-1} = 1 - i (a_o k)^{-1} - i r_o k + \mathcal{O}(k^3)\,. \ee
This result connects the strongly confined laboratory scenario to a scattering theory with a true 1D form and is consistent with previous calculations of p-wave scattering in strongly confined q1D systems ~\cite{Granger:2004im,Kanjilal:2004jv,Girardeau:2004,Pricoupenko:2008fq,Hess:2014hd,Gao:2015hz,Saeidian:2015db,Zhou:2017jv}. Effective scattering in the $m=0$ sector asymptotically resembles that of a spinless Fermi gas in 1D. 

Unitary odd-wave scattering, for which $f_o \to -1$, extends across the range of collision momenta  $|a_o|^{-1} \ll k \ll r_o^{-1}$. The effective range $r_o$ decreases with stronger transverse confinement, because it scales as $a_\perp^2/R_\mathrm{3D}$ and thus as $\omega_\perp^{-1}$. Experiments in this universal regime have not yet been realized, because a typical $k_F r_o$ is of order unity; thus the effective-range contribution must be kept. The persistent relevance of $r_0$ is due to the large ratio between the (optical) confinement length scale and the 3D effective range: $\lambda_L/R_\mathrm{3D}$ is approximately $300$. 
We also note that since the q1D approach uses a 3D interaction model at short range, our treatment would break down for $a_\perp \lesssim R_\mathrm{3D}$. The question of universality is discussed further in Sec.~\ref{ssec:width}. 
 
For atoms in the lowest even-wave scattering continuum, i.e. $\nrad=0$ and $m=\pm1$, or equivalently $0<\mathcal{E}_1 = (E_{\rm rel}-2\omega_\perp)/2\omega_\perp \ll 1$, the even-wave $T$-matrix, $T_{e}$, in the low-energy limit, reduces to
\begin{align}
&T_{e}^{-1} = \frac{2\mu }{2} \left[-a_e + k^2 r_e + \frac{i}{k}\right]\,, \quad \mbox{with} \nonumber \\
a_e &= -\frac{a_{\perp}}{2}\left[\frac{a_{\perp}^3}{6V_\mathrm{3D}} + \frac{2 a_{\perp}}{3 R_\mathrm{3D}} + \zeta(-1/2) + \zeta(1/2) \right] \nonumber \\
\mbox{and}\quad  r_e &= \frac{a_{\perp}^3}{16} \left[ \frac{4a_{\perp}}{3 R_\mathrm{3D}} + \zeta(1/2) + \zeta(3/2)\right]\,,
\label{eq:low_even}
\end{align}
where $a_e$ is the 1D even-wave scattering length and $r_e > 0$ is the 1D even-wave effective range. (See Tab.~\ref{tab:params} for a summary of scattering parameters.) This result is identical to a 1D calculation with true s-wave interactions~\cite{Olshanii:1998,Bergeman:2003kn,Moore:2004,Schmelcher:2005,Yurovsky:2005,Naidon:2007,Dunjko:2011,Kristensen:2015ib} with the effective range $r_e$ included. The corresponding 1D even-wave scattering amplitude is
\be \label{eq:fe1D}
-f_{e}^{-1} = 1 + i a_e k - i r_e k^3 + \mathcal{O}(k^5)\,. \ee
Compared to the odd-wave effective range term in Eq.~\eqref{eq:low_odd}, we see that the effective-range term, $k^3 r_e$ here, is smaller than $k r_o$ by a factor of $k^2 a_\perp^2$, 
which must be small in the q1D limit. As a result, both the resonant $k |a_e| \ll 1$ and zero-range $k^3 r_e \ll 1$ limits are achieved at low energy, with no necessary hierarchy between $a_e$ and $r_e$. At the largest lattice depth in our experiments, $k_F^3 r_e \sim 0.1$, which enables q1D even-wave scattering to enter a universal regime in which the effective range can be neglected. 

\subsection{RF spectrum \label{ssec:rf}}

The $T$-matrices \eqref{eq:low_odd} and \eqref{eq:low_even} allow us to calculate the rf lineshape, because at high frequency, the rf spectrum is related to the short-range correlations in the many-body system. Our starting point is the relation between $\Gamma$ and the 3D correlation strength \cite{Braaten:2008tc,Son:2010}:
\begin{align}
\Gamma(\omega) &= \frac{\Omega^2}{4} {\rm Im} \ i \! \int dt \, e^{i(\omega+i \varepsilon)t} \int d^3 \! {\bm R} \int d^3 {\bm r} \nonumber \\
&\left\langle T_t \ \hat\psi_b^{\dagger}\hat\psi_c\left({\bm R} + \frac{1}{2}{\bm r}, t\right)\hat\psi_c^{\dagger}\hat\psi_b\left({\bm R} - \frac{1}{2}{\bm r}, 0\right) \right\rangle \,,
\label{eq:RF_def}
\end{align}
where $\Omega$ is the Rabi frequency, ${\bm r}$ and ${\bm R}$ are the relative and center-of-mass positions, $T_t$ is the time-ordering operator, and $\hat\psi_b$ and $\hat\psi_c$ are the fermionic annihilation operators for a fermion in hyperfine-Zeeman state $\ket{b}$ and $\ket{c}$, respectively. Full expressions for the transfer rate are reported in App.~\ref{app:spectral}. In the experimentally relevant regime, where $\omega \gg E_F$ and $\omega < 4\omega_\perp$, 
\be \label{eq:Gamma02}
\Gamma^{(0)}(\omega) = \frac{\Omega^2C_o}{2}\left[\frac{\theta(\omega)}{\omega^{3/2}}+\frac{\theta(\omega-2\omega_\perp) \sqrt{\omega-2\omega_\perp}}{\omega^2}\right] \ee
and 
\be \label{eq:Gamma12}
\Gamma^{(\pm 1)}(\omega) =\frac{\Omega^2 C^{(\pm 1)}_e}{2}\left[\frac{\theta(\omega)}{\omega^{5/2}}+\frac{2\theta(\omega-2\omega_\perp)}{(\omega-2\omega_\perp)^{1/2}\omega^2}\right]\, . \ee
where $C_o$ is the 1D odd-wave contact and $C^{(\pm 1)}_e$ are the 1D even-wave contacts and $\theta(x)$ is the Heaviside step function. In both expressions, the first term represents the contribution from those continuum states that are below the first orbital gap. The step function $\theta(\omega-2\omega_{\perp})$ in the second terms indicate the opening of a new transverse band, when the rf photon energy exceeds the energy difference to the $\nrad=1$ scattering continuum, thus making the final scattering states within the new transverse band available. Each continuum has a density of states that scales as $1/\sqrt{\mathcal{E}_m}$, as is characteristic of a 1D system. 
This gives rise to a cusp-like $(\omega-2\omega_{\perp})^{1/2}$ dependence for $\Gamma^{(0)}$, and a singular dependence $(\omega-2\omega_{\perp})^{-1/2}$ for $\Gamma^{(\pm 1)}(\omega)$. The latter is observed experimentally: see Sec.~\ref{ssec:evenspec}. 

As discussed in App.~\ref{app:spectral}, both Eq.~\eqref{eq:Gamma02} and Eq.~\eqref{eq:Gamma12} are constrained to match the odd-wave contact relation with a characteristic $\omega^{-3/2}$ spin-flip rate for $\omega<2\omega_\perp$~\cite{Bender:2005gka,He:2016gh,Cui:2016kd,Cui:2016ho,Sekino:2018gk,Yin:2018hl,Maki:2021}, 
\be \label{eq:GamOdd1D}
\lim_{E_F/\hbar \ll \omega < 2 \omega_\perp } \Gamma^{(0)}(\omega) = \frac{\Omega^2 }2
\sqrt{\frac{2 \mu}{\hbar}} 
\frac{C_o}{\omega^{3/2}} 
\ee
and the even-wave contact relation with a leading $\omega^{-5/2}$ rate,
\be \label{eq:GamEven1D}
\lim_{E_F/\hbar \ll \omega < 2 \omega_\perp } \Gamma^{(\pm 1)}(\omega) = \frac{\Omega^2}2 \frac{1}{\sqrt{2 \hbar \mu}} \frac{C^{(\pm 1)}_e}{\omega^{5/2}}\,,
\ee
where factors of $\hbar$ have now been restored. These relations will be used to find the contact parameters in the measured spectra. 

\begin{figure*}[t!]
\centering
\includegraphics[width=2\columnwidth]{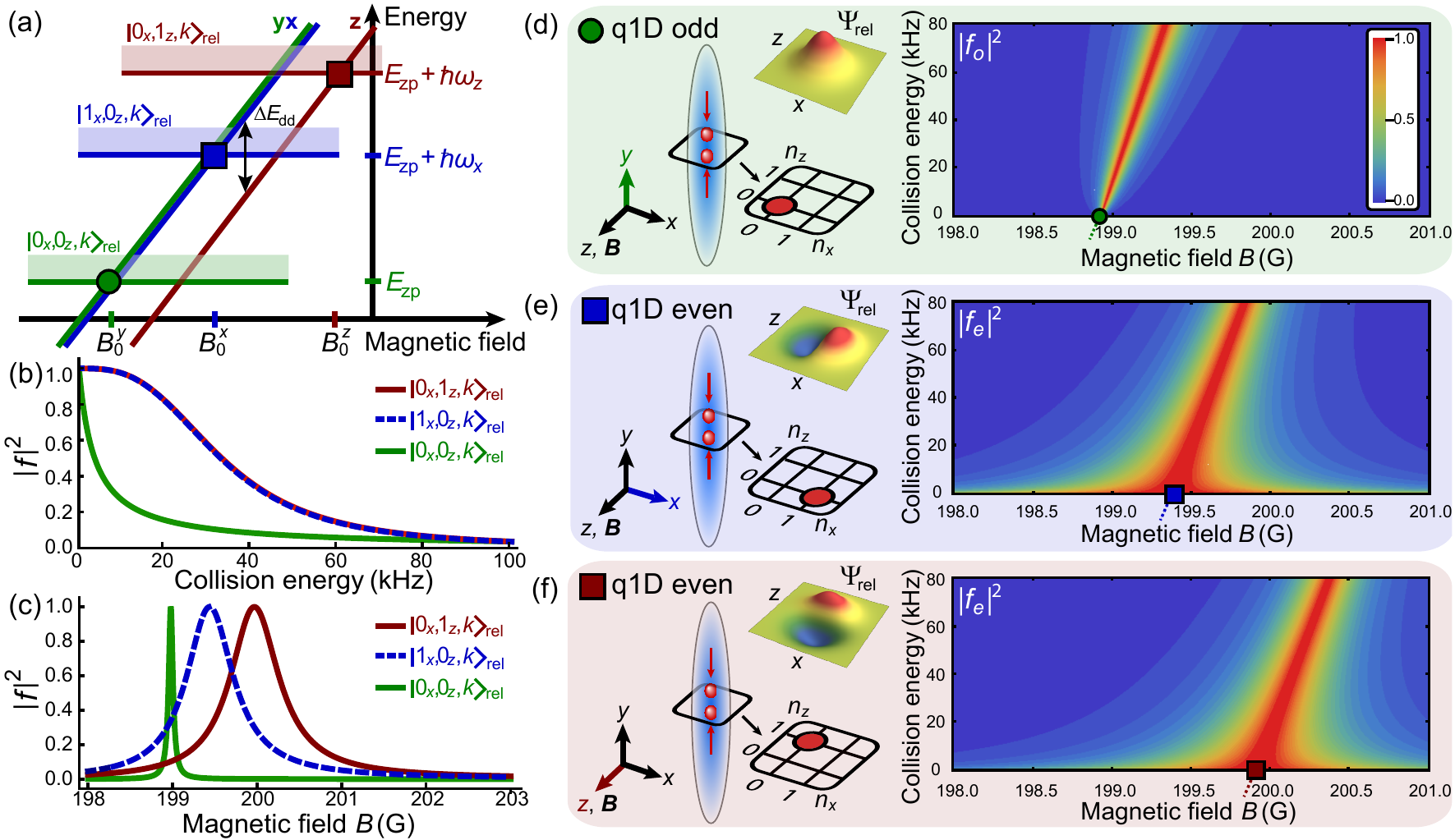}
\caption{
{\em Scattering resonances in q1D.} 
(a) The asymptotic states $\ket{n_x,n_z,k}_\mathrm{rel}$ have confinement energies of $(n_x+1/2)\hbar \omega_x + (n_z+1/2)\hbar \omega_z$ plus a continuum of $\hbar^2 k^2/2\mu$ from motion along $y$. The three lowest continua, shown here, become strongly interacting when magnetically tuned to resonance with an appropriate 3D p-wave dimer state (labeled $x$, $y$, and $z$). 
The resonant location for $\ket{1_x, 0_z, k}_\mathrm{rel}$ states (blue square at $B_0^x$) is at an energy shifted above the ground-band by the transverse excitation energy $\hbar \omega_x$. 
The resonance for $\ket{0_x, 1_z, k}_\mathrm{rel}$ states (red square at $B_0^z$), is further displaced by the dipole-dipole splitting $\Delta E_{\mathrm{dd}}$ of the $xy$ and $z$ dimer states, and any cylindrical asymmetry for which the case $\omega_z > \omega_x$ is shown. Corrections to this approximate picture are given by Eqs.~\eqref{eq:B0odd} and \eqref{eq:B0even}. 
(b) The effective 1D scattering cross section $|f|^2$ at resonance versus q1D collision energy. The broader resonance for asymptotic states with $n_x=1$ or $n_z=1$ states can be attributed to their even-wave character, as discussed in the text. 
(c) $|f|^2$ versus magnetic field, for a typical in-tube scattering energy of $10\,$kHz. 
(d) The q1D odd-wave continuum is characterized by a transverse wave function $\Psi_{\mathrm{rel}}(x,z)$ with even spatial parity. A contour plot shows the $|f_o|^2$ versus magnetic field and collision energy. 
(e, f) The q1D even-wave continua are each characterized by an asymptotic transverse wave function $\Psi_{\mathrm{rel}}(x,z)$ with odd spatial parity. 
The contour plots of $|f_e|^2$ versus magnetic field and collision energy combines the trends of panels (b) and (c), and emphasize the growing magnetic width in the $k \to 0$ limit that is characteristic of s-wave collisions.
Scattering amplitudes in panels (b) to (f) are based on Eq.~\eqref{eq:fo1D} with parameters in Eq.~\eqref{eq:low_odd}, or Eq.~\eqref{eq:fe1D} with parameters in Eq.~\eqref{eq:low_even}, with equal $V_L^x=V_L^z = 40\,E_R$.}
\label{fig:scattering1D} 
\end{figure*}

\subsection{Anisotropic interactions \label{ssec:anisotropy}}

\begin{table}[b!]
  \caption{Exchange parity and spatial orientation of the two-atom states in the lowest three scattering continua.
  \label{tab:parity}}
\begin{ruledtabular}
\begin{tabular}{ccccc}
   \multirow{2}{*}{\centering continuum}  & asymptotic & transverse & q1D & resonant\\ 
     & state & parity & parity & dimer\\ \hline
    \mbox{zeroth} & $\ket{0_x, 0_z, k}_\mathrm{rel}$ & even & odd & $y$ dimer\\ 
    \mbox{first} & $\ket{1_x, 0_z, k}_\mathrm{rel}$ & odd & even & $x$ dimer\\
    \mbox{second} & $\ket{0_x, 1_z, k}_\mathrm{rel}$ & odd & even & $z$ dimer\\ 
  \end{tabular}
  \end{ruledtabular}
\end{table}

Our treatment so far has assumed isotropic 3D interactions and radially symmetric trapping potentials. For the experimental system, however, interactions are anisotropic: the 3D $xy$ dimer branch is split by $\Delta E_{\mathrm{dd}}/h\sim100$\,kHz from the 3D $z$ dimer branch due to dipole-dipole interactions in $^{40}$K  \cite{Ticknor:2004}, leading to direction-dependent scattering volumes $V_{\mathrm{3D}}^{i}$ and effective ranges $R_{\textrm{3D}}^{i}$, where $i$ = $x$, $y$, or $z$ [see Fig.~\ref{fig:methods}(a) for axis orientation]. Furthermore the cylindrical anisotropy can be tuned by adjusting confinement strength independently along the $x$ and $z$ axes, as depicted by Fig.~\ref{fig:scattering1D}(a) in the case of $\omega_z > \omega_x$. 

For broken cylindrical symmetry, $\nrad$ and $m$ are no longer good quantum numbers. Instead, we use $n_x$ and $n_z$ to index quanta of relative motion in the Cartesian axes of confinement \footnote{The cylindrically symmetric states are superpositions of Cartesian states with $n_x + n_z = 2 \nrad + |m|$} and write the states as $\ket{n_x, n_z, k}_{\mathrm{rel}}$. Even parity in $y$ occurs for odd $n_x + n_z$, and vice versa. The exchange antisymmetry of the $\ket{1_x,0_z}_\mathrm{rel}$ and $\ket{0_x,1_z}_\mathrm{rel}$ states can be seen by decomposing them back into a single-particle basis. Labeling the colliding particles A and B, a $\ket{1}_\mathrm{rel}$ state with no CM excitation is $(\ket{1}_\mathrm{A} \ket{0}_\mathrm{B} - \ket{0}_\mathrm{A} \ket{1}_\mathrm{B})/\sqrt{2}$, or in other words, a singlet-like state in which two particles share a single quantum of orbital excitation. 

The emergence of q1D resonances is due to the hybridization of the 1D bands of scattering states with p-wave dimers that are now labeled as $x$-, $y$- and $z$-dimers due to dipole splitting [see Tab.~\ref{tab:parity} and Fig.~\ref{fig:scattering1D}(a)]. Due to the symmetry of the 3D dimers, a selection rule arises regarding the coupling to the scattering states. For example, the $x$-dimer has the short-range symmetry corresponding to the spherical harmonic $Y_{1x}\sim x/r$. As a result, the $x$ dimer can only be hybridized with continuum states in the $\ket{1_x,0_z,k}_\mathrm{rel}$ band above the threshold energy $E_{\mathrm{zp}} + \hbar \omega_x$, where $E_{\mathrm{zp}}$ is the zero-point energy $\frac12\hbar (\omega_x + \omega_z)$. The magnetic field at which this crossing occurs is labeled $B_0^{x}$ in Fig.~\ref{fig:scattering1D}(a). Similar analyses also apply to the $y$ and $z$ scattering: the $y$-dimer enhances  odd-wave scattering within the $\ket{0_x,0_z,k}_\mathrm{rel}$ band, and the $z$-dimer enhances  even-wave scattering in the $\ket{0_x,1_z,k}_\mathrm{rel}$ band (see Tab.~\ref{tab:parity}). 
The three distinct resonances are shown in Fig.~\ref{fig:scattering1D}(a) and are labeled in order of increasing magnetic field: $B_0^{y}$ (``zeroth''), $B_0^{x}$ (``first''), and $B_0^{z}$ (``second''). The nature of the scattering state is illustrated in Figs.~\ref{fig:scattering1D}(d), \ref{fig:scattering1D}(e), and \ref{fig:scattering1D}(f). 

In the presence of anisotropic interactions, we still calculate the q1D scattering parameters using Eqs.~\eqref{eq:low_odd} and \eqref{eq:low_even}, but now replace $V_\mathrm{3D}$ and $R_\mathrm{3D}$ with the values of $V_\mathrm{3D}^{i}$ and $R_\mathrm{3D}^{i}$ specific to the underlying 3D resonance \cite{AhmedBraun:2021}. The low-energy scattering parameters in Eq.~\eqref{eq:low_odd} and Eq.~\eqref{eq:low_even} can then be tuned by changing the 3D scattering parameters in order to engineer arbitrarily strong odd- or even-wave interactions.

In terms of the 3D quasi-bound-state energy \cite{Ticknor:2004,AhmedBraun:2021}, approximately $-\hbar^2 R_\mathrm{3D}^{i}/(2\mu V_{\mathrm{3D}}^{i})$, the resonance condition $a_o^{-1} = 0$ for odd-wave scattering is, from Eq.~\eqref{eq:fo1D}
\be \label{eq:B0odd}
- \frac{\hbar^2R_\mathrm{3D}^{y}}{2\mu V_\mathrm{3D}^{y}} = E_{\textrm{zp}} \left[ 1 - 6 \frac{R_\mathrm{3D}^{y}}{a_{\perp}}\zeta(-1/2) \right]\,,
\ee
while for even-wave scattering, the resonant conditions ($a_e^x = 0$ or $a_e^z = 0$) are, from Eq.~\eqref{eq:fe1D}
\begin{align}\label{eq:B0even}
- \frac{\hbar^2R_\mathrm{3D}^{x,z}}{2\mu V_\mathrm{3D}^{x,z}} = & (E_{\textrm{zp}} + \hbar \omega_{x,z}) \nonumber \\
&\times \left\{ 1 + \frac{3}{2} \frac{R_\mathrm{3D}^{x,z}}{a_{\perp}}\left[\zeta(1/2) + \zeta(-1/2)\right] \right\} \,.
\end{align}
Since $R^i_\mathrm{3D}/a_{\perp}$ is typically small in our experiments, these relations show that resonant odd- and even-wave scattering occurs approximately when the 3D quasi-bound-state reaches the relevant scattering continuum, as depicted in Fig.~\ref{fig:scattering1D}(a). 

The interaction anisotropy is observed in radio-frequency spectroscopy and thus the contact parameters. We note that the odd-wave contact relation is only governed by the interactions along the $y$-direction, so Eq.~\eqref{eq:Gamma02} is not altered. On the other hand, even-wave interactions now appear as $x$ and $z$ correlation peaks at distinct magnetic-field values, instead of the degenerate $m=\pm1$ correlation peaks in Eq.~\eqref{eq:Gamma12}. Asymptotic rf transfer rates are now proportional to $C_{e}^{x}$ and $C_{e}^{z}$, respectively. 

\subsection{Width of the universal regime \label{ssec:width}}

The qualitative departure of even-wave scattering from odd-wave is illustrated from its energetic width, i.e., the scaling of elastic scattering versus momentum $k$ of the q1D collision. 
Figures~\ref{fig:scattering1D}(b)--(f) show the scattering probabilities $|f_o|^2$ and $|f_e|^2$, as a function of both energy and magnetic field. A key feature is the enhanced width, in either parameter, of $|f_e|^2$. Near resonance, the energy range of unitary scattering is determined by the effective-range term of Eqs.~\eqref{eq:fo1D} and \eqref{eq:fe1D}. For odd-wave interactions, this energetic width can be characterized by $\Delta_o \equiv (\hbar^2/2\mu) r_o^{-2}$, while for even-wave interactions the energetic width is $\Delta_e \equiv (\hbar^2/2\mu) r_e^{-2/3}$. From Eqs.~\eqref{eq:low_odd} and \eqref{eq:low_even}, we find that $\Delta_o/\Delta_e = 18^{2/3} (R_\mathrm{3D}/a_\perp)^{4/3}$, which is small in experiments. This behavior is illustrated in Fig.~\ref{fig:scattering1D}(b) on resonance, which is a vertical cut at the resonant magnetic field in the contour plots of $|f_{o,e}|^2$ seen in panels (d), (e), and (f). 

If $a_e$ is tuned to zero [for instance at $B=B_0^{z}$ defined by Eq.~\eqref{eq:B0even}], then the unitary limit $f_e \to -1$ is achieved for $r_e k^3 \ll 1$. The range of scattering energies in which this occurs is given by $\hbar^2 k^2/2\mu \lesssim \Delta_e$. In terms of transverse confinement strength, the leading-order width is
\footnote{The sub-leading correction modifies $\Delta_e$ by $1 - 2 R_\mathrm{3D} a_\perp^{-1} [\zeta(1/2) + \zeta(3/2)]$, such that the leading order gives roughly a 10\% underestimate for our typical parameters.}
\be \Delta_e \approx \hbar \omega_\perp (3 \sqrt{2} R_{\mathrm 3D}/a_\perp)^{2/3}\,. \ee
Within a degenerate Fermi gas, the range of two-body collision energies is $2 E_F$. Scattering is unitary when the energetic width $\Delta_e \gtrsim 2\,E_F$, or
\be \label{eq:nle} 
N \lambda \lesssim (3 R_\mathrm{3D} /2 a_\perp)^{2/3}\,. \ee
The r.h.s.\ of Eq.~\eqref{eq:nle} is $\sim 0.2$ for the confinement strength typically used in low-dimensional quantum gas experiments, and thus poses a comparable constraint to the q1D criterion $N \lambda \ll 1$. For the $N \lambda \sim 0.1$ conditions in our experiments, resonant even-wave scattering is already energetically broad [see Fig.~\ref{fig:scattering1D}(b)]. 

Elastic odd-wave collisions also have a zero-range limit. A parallel argument finds that when $|a_o|^{-1}\to 0$, the energetic width of the unitary regime ($\Delta_o = \hbar^2 /2\mu r_o^{2}$) is $\Delta_o \approx \hbar \omega_\perp (18 R_\mathrm{3D}^2/a_\perp^2)$ to leading order 
\footnote{The sub-leading correction modifies $\Delta_o$ by $1 - 3 R_\mathrm{3D}\zeta(1/2)/a_\perp$, such that the leading order gives roughly a 20\% underestimate for our typical parameters.}. 
Comparing this width to $2\,E_F$, one finds that the confinement condition for ${ \Delta_o} \gtrsim 2\,E_F$ is \be \label{eq:nlo}
N \lambda \lesssim (3 R_\mathrm{3D} / a_\perp)^2\,. \ee
The r.h.s.\ of Eq.~\eqref{eq:nlo} is $\sim 0.02$ for our typical conditions, so is much more restrictive than the q1D criterion, and not satisfied for our experiments. 

The {\em magnetic} width of low-energy resonant scattering reveals qualitatively different scaling in the odd- and even-wave cases. From Eq.~\eqref{eq:fo1D}, at constant $a_o$, as $k \to 0$, we see that $|f_o|^2 \to 0$ and scattering is thus progressively narrower in $B$ at lower energy [Fig.~\ref{fig:scattering1D}(d)]. A similar analysis for Eq.~\eqref{eq:fe1D}, however, shows that $f_e \to -1$ for a progressively broader range in $B$ at lower energy [Figs.~\ref{fig:scattering1D}(e) and \ref{fig:scattering1D}(f)]. At a constant scattering energy, both odd- and even-wave resonances become broader in field with increasing confinement. However, the odd-wave field width scales with the square root of the lattice depth, while the even-wave field width scales linearly with lattice depth. Therefore, confinement can lead to a significantly broadened resonance for even-wave scattering, even though the energetic width of the underlying 3D resonance is still narrow 
\footnote{This can be seen from Eqs.~\eqref{eq:low_odd} and \eqref{eq:low_even} 
since the 3D scattering volume scales as $V_\mathrm{3D} \sim 1/B$ near the p-wave Feshbach resonance. The widening field range of correlations at higher lattice depths can be seen in Fig.~\ref{fig:results}, with the effect strongest for even-wave correlations in Fig.~\ref{fig:results}(b). There, even-wave correlations extend across a 3\,G range at $V_L=120\,E_R$, compared to a $\lesssim 0.5\,$G range of odd-wave correlations at similar lattice depth. Both of these are larger than the $\sim0.2$\,G range of significant correlations observed in 3D.}. 

In the zero-range limit identified here, the emergent even-wave scattering channel has a solitary relevant parameter that is tunable. This provides a regime in which a spin-polarized Fermi gas can be considered universal. Without the need for effective-range terms, its behavior can be understood with models independent of the particular element used in experiments, similar to 3D fermionic spin mixtures near an s-wave scattering resonance or two-dimensional bosons near their non-interacting fixed point \cite{ZwergerBook,Bloch:2008gl}.

\section{Observation of odd-wave correlations \label{sec:oddExp}}

\begin{figure}[tb!]
\centering
\includegraphics[width=\columnwidth]{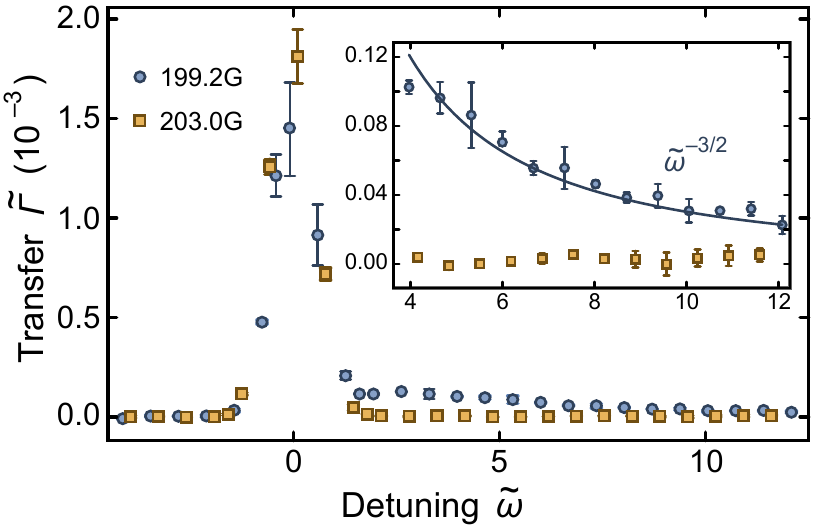}
\caption{{\em RF spectrum of odd-wave interactions.} 
The normalized rf transfer rate $\widetilde\Gamma$ versus dimensionless detuning $\widetilde\omega$ from the single-particle resonance [see Fig.~\ref{fig:methods}(b,v)]. Two spectra are shown, both for $V_L=40\,E_R$: at $B=203.0\,$G (squares), the gas is weakly interacting; at $B=199.2$\,G (circles), odd-wave interactions are Feshbach-enhanced [$-1/k_F a_o \sim 5$; see Fig.~\ref{fig:results}(a)]. 
The interacting gas shows an increased $\widetilde\Gamma$ for positive $\widetilde\omega$. 
The inset shows $\widetilde\Gamma(\widetilde\omega)$ on the same axes, with a $\widetilde\omega^{-3/2}$ fit to the high-frequency tail of the interacting spectrum, allowing the determination of $C_o$ from Eq.~\eqref{eq:Gamma02}. 
\label{fig:oddspectrum}}
\end{figure}

We probe odd-wave interactions by performing rf spectroscopy at magnetic fields near $B^y_0$ using the procedure outlined in Sec.~\ref{sec:exp}. Figure~\ref{fig:oddspectrum} shows the dimensionless transfer fraction $\widetilde{\Gamma}(\widetilde\omega)$ versus dimensionless rf detuning $\widetilde{\omega}\equiv \hbar \omega/E_F$, both rescaled by the calculated Fermi energy of the central tube (see App.~\ref{app:inhomogeneity}). Correlations are revealed by 
an asymmetric spectrum, i.e., spectral weight that appears at a large positive $\omega$ but not at a large negative $\omega$ of equal magnitude. 
The strength of correlations can be quantified through Eq.~\eqref{eq:GamOdd1D}, in the range $4 \lesssim \widetilde{\omega} \lesssim 15$ (for which $\omega < 2\omega_\perp$), where we observed the characteristic $\widetilde{\omega}^{-3/2}$ power-law decrease. 
In the strongly interacting regime, we fit the experimental data to extract $C_o$, thereby measuring the odd-wave correlations in a q1D ultracold gas for the first time to our knowledge. 
At magnetic fields far from resonance, correlations are weak and there is no observable asymmetry, as can be seen in the inset of Fig.~\ref{fig:oddspectrum}.

\begin{figure}[tb!]
\centering
\includegraphics[width=\columnwidth]{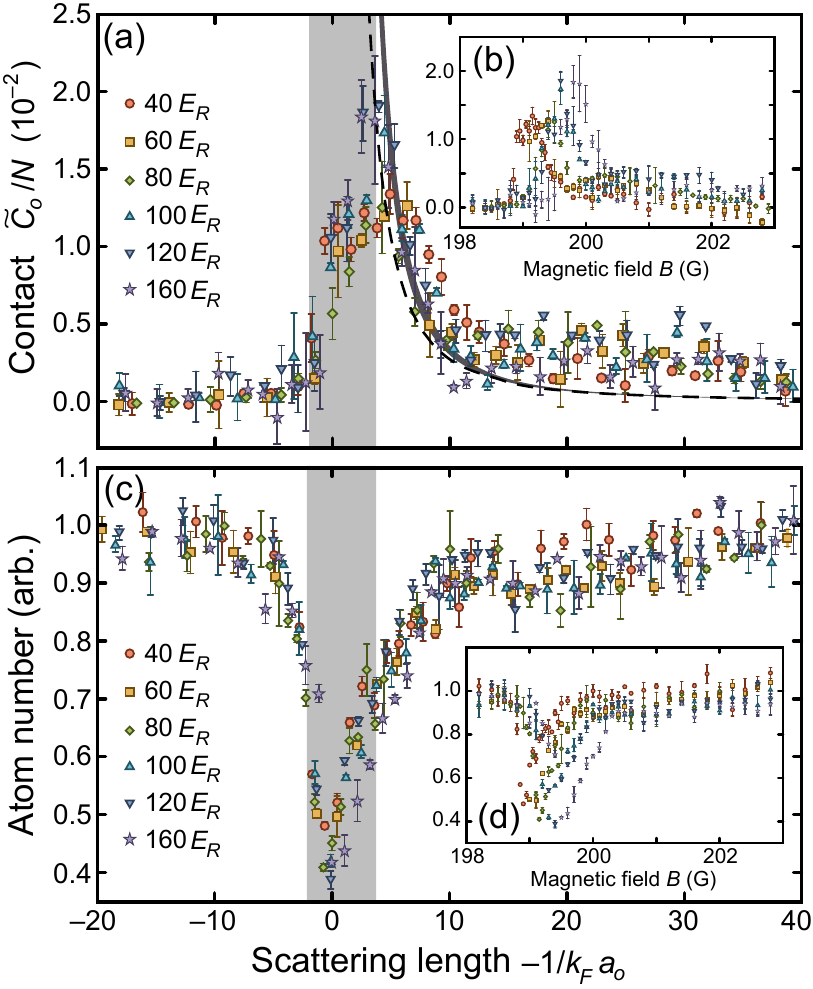}
\caption{{\em Odd-wave contact and loss.} 
(a) Contact per atom versus odd-wave scattering length, $-1/k_Fa_o$. Per Eq.~\eqref{eq:rf}, the dimensionless odd-wave contact per atom is related to the scaled transfer from spectroscopy.
The zero-temperature Bethe ansatz prediction for $k_F r_o = 1$ is shown in dashed-black, and the high-temperature prediction for $T/T_F=0.8$ is shown as a band covering the range of experimental lattice depths. (b) The inset displays the same $\widetilde{C}_o/N$ data versus magnetic field, emphasizing the data collapse in (a).
(c) The total atom number after the spectroscopy sequence, as a fraction of initial atom number, is shown as a function of $-1/k_Fa_o$. In the grey-shaded parameter range, 30\% or more atoms are lost. Peak loss occurs at the calculated $a_o^{-1}=0$. 
(d) The inset displays this atom-number data as a function of magnetic field.
}
\label{fig:oddcontact} 
\end{figure}

Figure~\ref{fig:oddcontact} shows the contact strength and atom number measured across a range of magnetic fields and lattice depths. The contact was measured using transfer fraction at a single frequency, typically $\widetilde{\omega}=5$, which fulfills the condition $E_F/\hbar \ll \omega < 2 \omega_\perp$ and selects the first term of Eq.~\eqref{eq:Gamma02}. Since we are optimizing for ground-band loading, we can assume that $\Gamma^{(\pm 1)} \ll \Gamma^{(0)}$, which allows for a single-shot determination of contact:
\be \label{eq:rf} 
\frac{\widetilde{C}_o}{N} = 4 \sqrt{2} \pi \, \widetilde\omega^{3/2} 
\, \widetilde\Gamma \,, \ee 
where 
\be \widetilde{C}_o \equiv \frac{4 m}{\hbar^2 k_F} C_o \ee 
is the dimensionless odd-wave contact 
\footnote{The numerical factors in $\tilde{C}$ are chosen such that the asymptotic momentum density $n_k \to \tilde{C}_o k_F/k^2$, when $\int (dk/2 \pi) n_k = N$.}. 

The observed magnitude of $\widetilde{C}_o/N$ is $\lesssim 2 \times 10^{-2}$, with the maximum near the q1D scattering resonance. In Figs.~\ref{fig:oddcontact}(b,d), both atom loss and correlations shift to higher magnetic field with increasing lattice depth, primarily due to the increasing zero-point energy, $E_{\textrm{zp}}$, associated with the transverse harmonic confinement. When comparing Fig.~\ref{fig:oddcontact}(a) to Fig.~\ref{fig:oddcontact}(b), or comparing Fig.~\ref{fig:oddcontact}(c) to Fig.~\ref{fig:oddcontact}(d), we see that field-dependent data collapses to a single curve when plotted as a function of $-1/k_Fa_o$, indicating that we are probing the physics in the q1D regime. 

Consider first the weakly repulsive regime. No significant correlations are seen in the range $-20 < -1/k_F a_o < -2$, which can be understood from Fig.~\ref{fig:scattering1D}(a): the 3D quasi-bound p-wave dimer has yet to emerge above the odd-wave scattering threshold. 
Similar to the 3D p-wave case \cite{Luciuk:2016gr}, we conclude that the dimer state is not significantly populated. For stronger repulsive interactions \footnote{Note that inhomogeneous transverse confinement broadens the range of $a_o^{-1}$ for a given magnetic field, which makes the assignment of small $-1/k_F a_o$ ambiguous.},
$-5 \lesssim -1/k_F a_o < 0$, a finite $\Tilde{C}_o$ is observed, however concurrently with strong losses, which are perhaps due to the instability of the repulsive odd-wave gas \cite{Pan:2018}, its value remains small. 

Next, consider the regime with relatively weak attractive interactions, $4 \lesssim -1/k_F a_o \lesssim 15$. In the absence of strong loss, we can compare the measured values of $C_o$ to equilibrium calculations: zero-temperature Bethe ansatz theory \cite{Yin:2018hl} and a high-temperature expansion theory \cite{Maki:2021} (App.~\ref{app:hte}). These are shown in Fig.~\ref{fig:oddcontact}(a). 
The theory band for the high-temperature expansion reflects the weak 1D effective-range dependence on lattice depth and includes a correction for the inhomogeneous density and tube number discussed in App.~\ref{app:inhomogeneity}. In this weakly interacting regime, the measurements overlap with both the zero-temperature and the high-temperature predictions, with no adjustable parameters, and follow the anticipated monotonic decrease with $a_o^{-1}$. 

At higher magnetic fields, where $-1/k_F a_o \gtrsim 15$, the apparent contact deviates from these predictions. We attribute this departure to overlapping even-wave correlations (see Sec.~\ref{sec:evenExp}), due to a remaining population in the excited bands, such that $\Gamma$ can no longer be associated purely with $\Gamma^{(0)}$. 

Finally, consider magnetic fields close to resonance ($B \sim B^y_0$), where we observe a smaller contact than is anticipated by theory. The unitary high-temperature value of the contact is 
$\widetilde{C}_o/N = 8/\pi$ in a uniform system \cite{Maki:2021} while at zero temperature $\widetilde{C}_o/N = 4/\pi$ \cite{Bender:2005gka,Cui:2016kd,Sekino:2018gk}. 
This far exceeds the highest $\widetilde{C}_o/N$ observed. 
Instead, within the range $-2 \lesssim -1/k_F a_o \lesssim 4$, indicated in Fig.~\ref{fig:oddcontact}(c) with a gray band, the contact per atom is decreased. This is attributed to the onset of strong loss: we find that more than 30\% of the initial population is lost by the end of $t_\mathrm{hold}$, from which we infer a three-body loss coefficient $L_3 \geq 10^{-5}$\,cm$^2$/s. As in prior studies of atom loss in q1D Fermi gases~\cite{Marcum:2020,Chang:2020}, we find that the loss rate is independent of lattice depth as indicated by data collapse in Fig.~\ref{fig:oddcontact}(c). 

Limited by losses, the system does not develop a contact parameter larger than $\widetilde{C}_o / N \sim 2 \times 10^{-2}$. We can interpret this magnitude by estimating the odd-wave interaction energy that would be found in an equilibrium cloud with the same correlation strength and negligible effective range. The adiabatic relation for 1D odd waves is \cite{Cui:2016kd}
\be \frac{\partial F}{\partial (-a_o^{-1})} = C_o \,, \ee
where $F$ is the free energy of the gas. Integrating under the observed contact outside of the loss dominated regime, we find that between $(-k_F a_o)^{-1} = 10$ and $(-k_F a_o)^{-1} = 4$, the interaction energy is $\Delta F \sim -0.03\,N E_F$. Therefore, at least in the conditions realized here, an odd-wave Fermi gas can be stable only in the weakly interacting regime. 

\section{Observation of even-wave correlations \label{sec:evenExp}}

We now turn our attention to the features that appear when multi-orbital loading is allowed, i.e., those features appearing in Fig.~\ref{fig:results}(b) but not in Fig.~\ref{fig:results}(a). In order to test the hypothesis that these spectral features are due to q1D even-wave correlations and to elucidate their nature, we undertake a sequence of experiments: spectroscopy with variable asymmetric confinement (Sec.~\ref{ssec:ani}); spectroscopy at variable frequency (Sec.~\ref{ssec:evenspec}); and spectroscopy with variable coupling strength (Sec.~\ref{ssec:ce}). The sample preparation common to all these tests does not use AM (see Sec.~\ref{sec:exp} and App.~\ref{app:bandpop}), such that $\sim 15\%$ of the atoms are in each of the transverse excited bands as measured by band mapping. 

\begin{figure}[tb!]
\centering
\includegraphics[width=1.0\columnwidth]{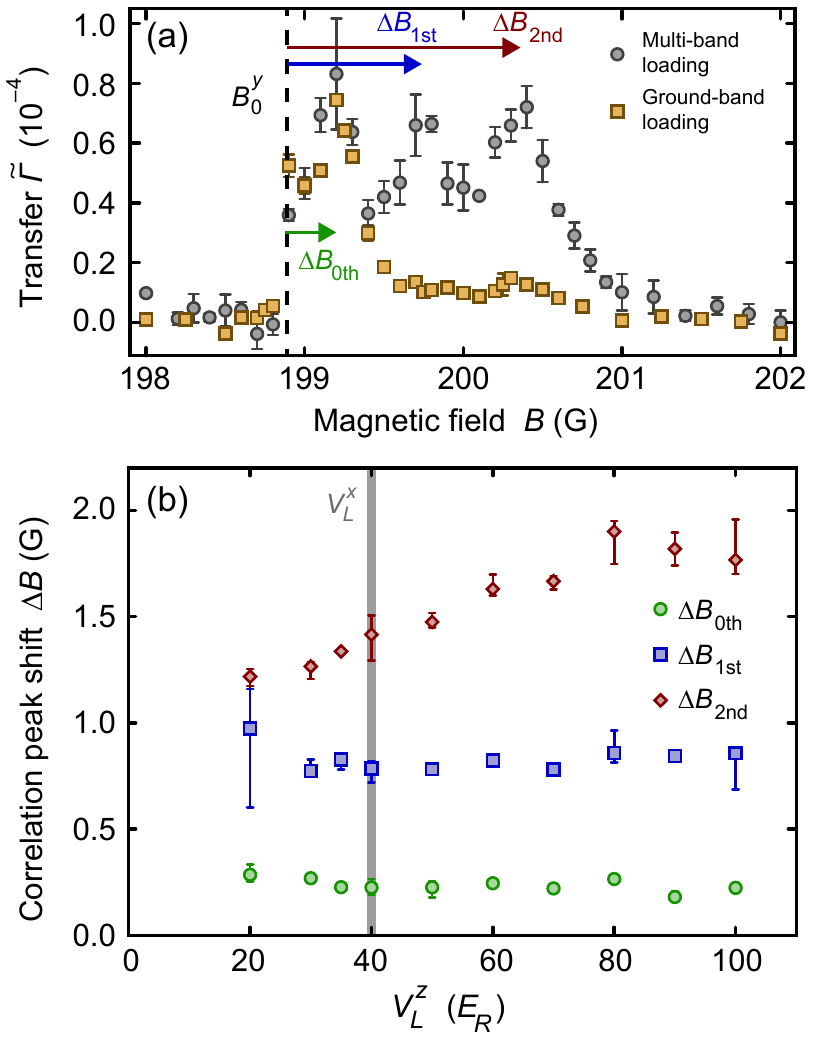}
\caption{{\em Vector assignment of the second even-wave correlation peak.} 
(a) Normalized rf transfer rate $\widetilde\Gamma$ at constant $\widetilde{\omega} \approx 5$ versus magnetic field $B$ applied in the $z$ direction. Three peaks in $\widetilde\Gamma(B)$ are resolved for multi-band loading (circles), while a single clear peak is observed for ground-band loading (squares). Here, confinement is cylindrically symmetric, with $V_{L}^{x}=V_{L}^{z}=40\,E_R$. The magnetic-field displacements from the resonance position $B_{0}^{y}$ are indicated for maximum $\widetilde\Gamma$ by colored arrows, and reported as $\Delta B$ in the subsequent panel. 
(b) The magnetic-field displacements for the three peaks in $\widetilde\Gamma(B)$ are shown for variable $V_{L}^{z}$, with fixed $V_{L}^{x}=40\,E_R$. While $B_\mathrm{0th}$ and $B_\mathrm{1st}$ shift with the zero-point energy included in $B_0^{y}$, $B_\mathrm{2nd}$ has a stronger dependence on $V_{L}^{z}$. This trend supports the hypothesis that the third resolved peak in $\widetilde\Gamma$ versus $B$ corresponds to scattering in the $z$ continuum, whose energy is $\hbar \omega_z$ above the odd-wave continuum. This coincides with the resonance condition for even-wave scattering, Eq.~\eqref{eq:approxB0xz}; more direct evidence of even-wave character is given in Sec.~\ref{ssec:evenspec}.
}
\label{fig:vector} 
\end{figure}

\subsection{Vector character of the even-wave resonances\label{ssec:ani}}

The resonance condition for even-wave interactions is given by Eq.~\eqref{eq:B0even}, which for our small value of $R_{\mathrm{3D}}/a_\perp$ can be approximated for the $i=x$ or $i=z$ continuua as
\be \label{eq:approxB0xz} 
- \left.\frac{\hbar^2R_\mathrm{3D}^i}{2\mu V_{\mathrm{3D}}^i}\right|_{B=B_0^i}
\approx E_\mathrm{zp} + \hbar \omega_i\,,\ee
where the l.h.s.\ is the 3D p-wave dimer energy, linear in $B$, as shown in Fig.~\ref{fig:scattering1D}(a).
The incorporation of an extra vibrational excitation quantum (shared between the atoms in a $\ket{1_x}_\mathrm{rel}$ or $\ket{1_z}_\mathrm{rel}$ motional state) shifts the even-wave resonances from the odd-wave resonance condition, which is 
\be - \left.\frac{\hbar^2R_\mathrm{3D}^{y}}{2\mu V_{\mathrm{3D}}^{y}}\right|_{B=B_0^y}
\approx E_\mathrm{zp}\, ,\ee 
from Eq.~\eqref{eq:B0odd}. In the following, we investigate the dependence of $B_0^i$ on $\omega_i$.

Figure~\ref{fig:vector}(a) shows $\widetilde\Gamma(B)$ at $V_L=40E_R$, a depth at which three distinct peaks can be resolved in multi-band loading. 
To subtract off the common shift from the zero-point energy, we show the displacements of each peak ($\Delta B_\mathrm{0th}$, $\Delta B_\mathrm{1st}$, and $\Delta B_\mathrm{2nd}$) from $B_0^y$
\footnote{The sub-leading term of the isotropic Eq.~\ref{eq:B0odd} 
provides only a $\sim 50\,$mG correction to the leading order. Still, we determine $B_0^y$ through a generalized form of Eq.~\eqref{eq:B0odd} that includes anisotropic confinement \cite{Gao:2015hz}.}.
Since $\Gamma = \sum \Gamma^i$, additional tests are necessary to confirm the origin of these peaks. 

The first peak also appears with a ground-band loading procedure [as shown in Fig.~\ref{fig:vector}(a) and Fig.~\ref{fig:results}(a)], so it can be unambiguously associated with odd-wave interactions. As shown in Fig.~\ref{fig:oddcontact}, the correlation peak is not at the resonance location, due to strong loss near resonance. For the parameters of Fig.~\ref{fig:vector}, the magnetic field location of peak correlation $B_\mathrm{0th}$ is shifted $\sim 0.2$\,G above $B_0^y$.

A vector assignment of these two higher peaks can be made using unequal confinement, $V_L^x \neq V_L^z$, whose anticipated effect on $B_0^{x}$ and $B_0^{z}$ is illustrated in Fig.~\ref{fig:scattering1D}(a) for $\omega_z > \omega_x$. 
Figure~\ref{fig:vector}(b) shows the best-fit peak displacements $\Delta B_\mathrm{0th}$, $\Delta B_\mathrm{1st}$, and $\Delta B_\mathrm{2nd}$ from repeated $\widetilde\Gamma(B)$ scans at variable $V_L^z$, with fixed $V_{L}^{x}=40\,E_R$. 
We see how $\Delta B_\mathrm{1st}$ is insensitive to $V_L^z$ (similarly to $\Delta B_\mathrm{0th}$), while $\Delta B_\mathrm{2nd}$ has a stronger dependence on $V_L^z$. Based on this trend, we can conclude that the peak at $B_\mathrm{2nd}$ must be associated with confinement along the $z$ direction, which would be expected for the even-wave threshold $B_0^z$, where scattering states hybridize with the 3D $z$-dimer.

Cylindrical symmetry is broken in our experiments by the applied magnetic field direction ($z$) orthogonal to the q1D direction ($y$). The field aligns the magnetic dipole moments of $^{40}$K, and thus splits the Feshbach resonances for the two even-wave features (see Sec.~\ref{ssec:anisotropy}). This $0.5$\,G splitting is evident upon the comparison of $B_\mathrm{1st}$ and $B_\mathrm{2nd}$ at equal lattice depth, which is $V_{L}^{z}=40\,E_R$ for Fig.~\ref{fig:vector}. 

A similar experimental study with variable $V_L^{x}$ and fixed $V_L^{z}$ indicates that the peak at $B_\mathrm{1st}$ is associated with confinement along $x$. In sum, the vector character and dependence on the lateral confinement is consistent with the assignment of the correlation peak at $B_\mathrm{1st}$ in Fig.~\ref{fig:vector} to the $x$ even-wave continuum, and at $B_\mathrm{2nd}$ to the $z$ even-wave continuum. 

\subsection{Spectroscopic signatures \label{ssec:evenspec}}

\begin{figure}[tb!]
\centering
\includegraphics[width=1\columnwidth]{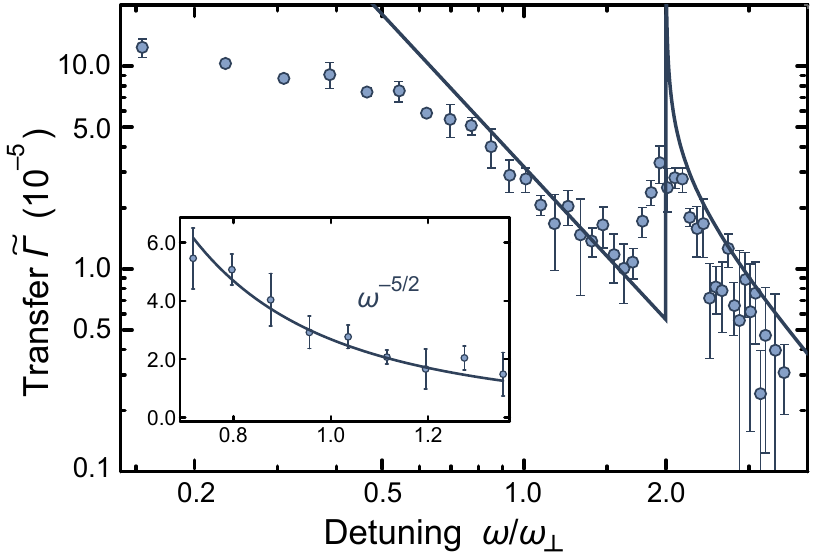}
\caption{{\em Even-wave rf spectrum.} 
Dimensionless rf transfer rate $\widetilde\Gamma$ versus detuning $\omega$, for a multi-band-loaded ensemble at 200.5\,G, with $V_L=80\,E_R$. Between $0.7\,\omega_\perp$ and $1.7\,\omega_\perp$, the signature even-wave scaling $\widetilde\Gamma \propto \omega^{-5/2}$ is observed [see Eq.~\eqref{eq:Contacts}]. 
The solid lines show Eq.~\eqref{eq:Gamma12}, with a free vertical scale factor. The singularity in the transfer rate (a unique signature of 1D even waves) manifests as a peak near $\omega \approx 2\omega_\perp$, as described in the main text. 
The inset shows $\widetilde\Gamma(\omega)$ on a linear vertical scale in the $\omega^{-5/2}$ regime; a single-parameter fit determines the sum of the even-wave contacts. For these conditions, $k_F a_e^z\approx 0.4$, compared to $k_F a_e^x \approx 1.6$, so that scattering is dominated by the $\ket{0_x, 1_z, k}_\mathrm{rel}$ states. 
}
\label{fig:evenspectrum} 
\end{figure}

The strongest evidence for the observation of even-wave correlations in our system comes from the functional form of the rf spectra. As discussed in Sec.~\ref{ssec:rf}, within the range of frequencies $\omega<2\omega_\perp$, in which a q1D system resembles a true 1D system, the rf transfer rate should scale as $\omega^{-5/2}$ (as opposed to $\omega^{-3/2}$ found for odd-wave correlations in Sec.~\ref{sec:oddExp}). Furthermore, for larger rf frequencies, $\widetilde{\Gamma}(\omega)$ should exhibit a singularity at $\omega=2\omega_\perp$: see Eq.~\eqref{eq:Gamma12}. 
The experimentally measured $\Gamma(\omega)$ is a sum of all $\Gamma^{i}$, but will be dominated by only one Cartesian component near its resonant magnetic field. 

Figure~\ref{fig:evenspectrum} shows the rf spectrum observed at $B\approx B_\mathrm{1st}$. For frequencies between $0.7\,\omega_\perp$ and $1.7\,\omega_\perp$, we observe a clear $-5/2$ power law consistent with 1D s-wave correlations (see Fig.~\ref{fig:evenspectrum} inset). In this region, it is possible to extract the 1D even wave contact parameters $C_e^{x,z}$. We note that once $C_e^{x}$ and $C_e^{z}$ are fixed, the general form of $\widetilde\Gamma(\omega)$ [Eq.~\eqref{eq:Gamma1} and in particular Eq.~\eqref{eq:Gamma12}] is then determined for higher frequencies, no matter what the relative strengths of $C_e^{x}$ and $C_e^{z}$. 
We do not understand why the $-5/2$ power law does not extend to lower frequencies, below $0.7\,\omega_\perp/2\pi \sim 90$\,kHz. This may be due to incomplete equilibration of the long-wavelength degrees of freedom   \cite{Makotyn:2014,Yin:2016,Eigen:2018}. 

A sharp peak in $\widetilde\Gamma(\omega)$ appears, as anticipated, at twice the band gap 
\footnote{Due to anharmonicity of the optical lattice potential, there is some ambiguity in the experimental value of the transverse trap frequency. Since the lowest allowed excitation for even-waves is the excitation of an atom from the first to third single-particle level of the optical lattice, the best choice for $\hbar \omega_\perp$ in this case is half that energy.}. %
Here, atom pairs are promoted from incident states with principal radial quantum number $\nrad=0$ to final states with $\nrad=1$ and hyperfine-Zeeman state $\ket{c}$. 
Since rf transitions cannot modify the spatial wave functions, the presence of the feature demonstrates that the scattering states are dressed by higher-$\nrad$ orbital states at short range, i.e., $y \lesssim a_\perp$ in Eq.~\eqref{eq:wf}. 
The singular feature in the theory results from a combination of two factors: the divergence of the 1D density of final states (see Sec.~\ref{ssec:rf}) and the $k$-independent even-wave matrix element [see the $|m|=1$ term of  Eq.~\eqref{eq:Umatrix}]. 
As observed, the $2 \omega_\perp$ singularity is softened by the distributions of a finite range of collision energies, the average over variable lattice depths of 1D tubes, and the Fourier broadening of the rf pulse. 

The observation of these two even-wave signatures in $\Gamma(\omega)$ gives strong evidence for the emergence of effective s-wave correlations in the q1D gas. 

\subsection{Even-wave contact parameter \label{ssec:ce} }

\begin{figure}[tb!]
\centering\includegraphics[width=1\columnwidth]{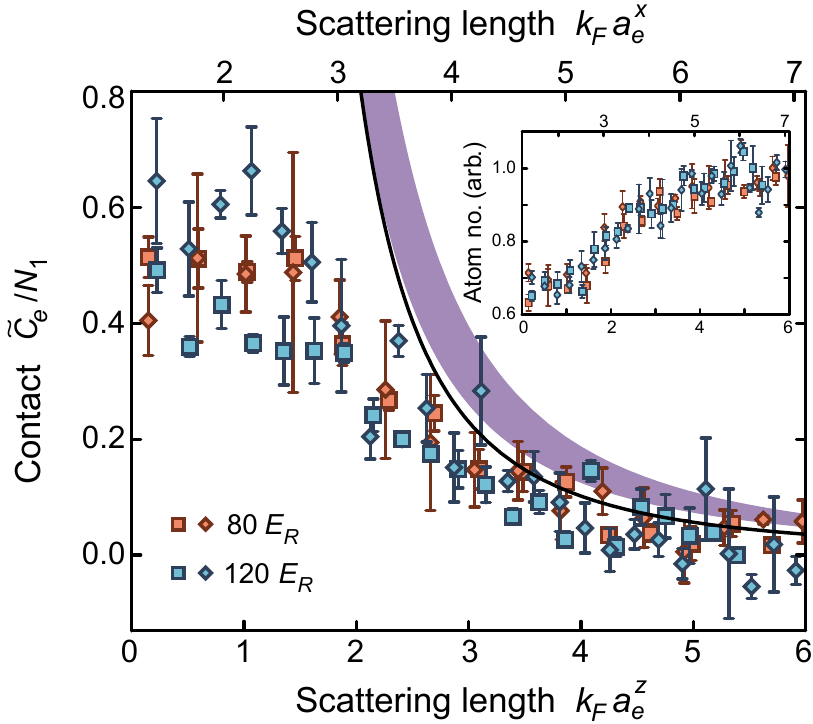}
\caption{{\em Even-wave correlations.} 
Even-wave contact $\widetilde C_e = \widetilde C_e^x + \widetilde C_e^z$ normalized by $N_1$, plotted versus even-wave scattering lengths $a_e^z$ (bottom) and $a_e^x$ (top). 
The scattering lengths $a_{e}$ are tuned by the magnetic field across a range 200\,G to 205\,G, and scaled by the calculated $k_F$ of the central tube. 
Single-shot measurements of $C_e$ use Eq.~\eqref{eq:even_singleshot}, repeated here for two different single-shot detunings: $\omega/2 \pi = 100$\,kHz (squares) and $\omega/2 \pi = 125$\,kHz (diamonds). Two lattice depths are shown: red and blue points are $80\,E_R$ and $120\,E_R$, respectively. 
The shaded purple band gives a high-temperature expansion prediction for $\widetilde{C}_e^{x}/N_1+\widetilde{C}_e^{z}/N_1$, from Eq.~\eqref{eq:Caint}, with no fit parameters. The width of the band reflects one standard deviation of systematic uncertainties (atom number and band-population fractions) and the variation of effective-range contributions. 
Both experiment and theory include both $x$ and $z$ even-wave interactions. However in the region of interest, correlations are dominated by even-wave $z$ pairing: a solid line shows $\widetilde{C}_e^{z}/N_1$ alone. The inset shows the relative atom number after spectroscopy, for the same ranges of $k_F a_e^z$ (bottom) and $k_F a_e^x$ (top). } 
\label{fig:evencontact} 
\end{figure}

The correlations appearing in Fig.~\ref{fig:results}(b) across a broad range in magnetic field can now be quantified through the even-wave contact. At each magnetic field, the even-wave contacts $C_e^{x,z}$ can be probed with a single-shot measurement of $\Gamma$ at a fixed detuning
with the range $0.7\,\omega_\perp \lesssim \omega \lesssim 1.7\,\omega_\perp$, as discussed in Sec.~\ref{ssec:evenspec}. In the regime where $\Gamma^{(0)} \ll \Gamma^{(x,z)}$, we find  
\be \label{eq:even_singleshot} 
\widetilde C_e^{x} + \widetilde C_e^{z}= 2 \sqrt{2} \pi N \widetilde{\omega}^{5/2} \widetilde \Gamma\,, \ee
where 
\be \label{eq:Catilde}
\widetilde{C}_e^{x,z} \equiv \frac{4m}{\hbar^2 k_F^3} C_e^{x,z} \ee 
is the dimensionless even-wave contact. 
Since only $\sim 15\%$ of atoms populate each transverse excited band, we base $k_F$ on the total atom number, just as in the odd-wave case. 
We show in App.~\ref{app:bandpop} that rf transfer is proportional to the excited-band population, motivating us to report the contact divided by the number of atoms in one single-particle excited motional excited band $N_1$, which we assume to be equal for both the $x$ and $z$ directions. In Fig.~\ref{fig:evencontact} we report $\widetilde{C}_e/N_1$ for various interaction strengths, tuning the scattering length $a_{e}^{z}$ across a range of $200\,$G to $205\,$G, away from the odd-wave feature. 
We observe data collapse between different lattice depths and single-shot detunings 
$\omega \sim \omega_\perp$ chosen to be within the $-5/2$ power law regime described in Sec.~\ref{ssec:evenspec}.

For $k_F a_e^z \gtrsim 2.5$, the observed trend matches that of a high-temperature model (App.~\ref{app:hte}) that maps the imbalanced band mixture in our system onto a spin mixture with true 1D s-wave interactions. 
For the large lattice depths in Fig.~\ref{fig:evencontact}, the distinct even-wave resonances overlap [see Fig.~\ref{fig:results}(b)] and the slightly larger $a_e^x$ is shown on the upper axis of Fig.~\ref{fig:evencontact}. Both experiment and theory include both $x$ and $z$ even-wave interactions. However in the region of interest, correlations are dominated by even-wave $z$ pairing as indicated by the agreement with a solid line that shows $\widetilde{C}_e^{z}/N_1$ alone. 

For $0 \leq k_F a_e^z \lesssim 2$, the observed $\widetilde{C}_e/N_1$ plateaus to $0.5(1)$. The expected near-unitary contact can be estimated using a zero-temperature theory \cite{He:2016gh} developed for 1D s-wave spin mixtures, again using a mapping of imbalanced band populations to imbalanced spin population in the model. In the homogeneous case, we find $\widetilde{C}_e^z/N_1\approx 1.0$, on the same order of magnitude as the observed value. A full description of the unitary regime would need to include loss [see inset of Fig.~\ref{fig:evencontact}], inhomogeneity, and the contribution of $C_e^x$. The contrast with the q1D odd-wave case, where the observed contact was $\lesssim 10^{-2}$ of the unitary value, is already striking. 

We can interpret the strength of the observed contact using the adiabatic relation from 1D s-waves \cite{Barth:2011}: 
\be \frac{\partial F}{\partial a_e^{z}} = C_e^{z}\,. \ee
Assuming thermal equilibrium, the interaction energy is then found by integrating under the observed contact, which contains contributions from both $x$ and $z$ resonances but is dominated by $\widetilde{C}_e^{z}$. As a lower estimate for the total interaction energy, $\Delta F$, we find that between $k_F a_e^{z} = 5$ and $k_F a_e^{z} = 2$, $ \sim -0.2\,N_1 E_F$, whereas integrating between $k_F a_e^{z} = 5$ and $k_F a_e^{z} = 0$ we find $\Delta F \sim -0.7\,N_1 E_F$. The q1D even-wave gas is thus strongly interacting, with an attractive energy that is at least an order of magnitude larger than is observed in the low-loss range of the odd-wave interactions. 

We can understand the enhanced strength of q1D even-wave interactions through the proportionality of p-wave coupling to momentum, as indicated by Eq.~\eqref{eq:Umatrix}. 
For odd waves, the relevant momentum that leads to a finite interaction is in the $y$ direction, along the tube. For even waves, p-wave coupling is achieved through the transverse gradient in the spatial wave function, which is proportional to $a_\perp^{-1}$ [see Eq.~\eqref{eq:int_q1d}]. 
For a typical momentum $k_F$, the ratio of even- to odd-wave coupling is thus 2/$a_\perp^{2} k_F^2$, which is of order $1/N \lambda$, and thus necessarily large in the q1D regime.  

\section{Discussion and Conclusion \label{sec:conclusion} }

The extension of low-dimensional systems to include multi-orbital scattering opens the door to several new prospects. We have shown that one result is the emergence of q1D even-wave interactions in a spin-polarized Fermi gas. The range of many-body physics accessible with this new interaction channel will be constrained by atom loss, including three-body recombination. Odd-wave atom loss processes have been studied both theoretically \cite{Levinsen:2008,Kurlov:2017,Sekino:2018gk} experimentally \cite{Chang:2020,Marcum:2020}; similar studies in the q1D even-wave scenario are called for, which also address band relaxation.

Future work could also include a more detailed understanding of rf spectra such as shown in Fig.~\ref{fig:oddspectrum} and Fig.~\ref{fig:evenspectrum}, especially the low-frequency deviations from the high-frequency power law. In this regime, the sub-leading contacts, conjugate to $r_o$ for odd waves \cite{Maki:2021} and $r_e$ for even waves, are expected to appear as additional higher-order power-law contributions to the rf spectral function $\Gamma$.

A shallow q1D dimer state lies below threshold both in the repulsive $a_o>0$ gas and in the attractive $a_e >0$ gas probed in our experiment. Searching for these dimer states and characterizing their stability would  further examine the paradigm of an emergent q1D symmetry. In the even-wave case, the binding energy is expected to be large even near resonance.

Another prospect is to search for emergent s-waves in quasi-two-dimensional (q2D) ensembles of identical fermions. A similar mechanism would apply: that antisymmetric orbital wave functions in a strong confinement direction would enable exchange-even collisional wave functions in the low-energy q2D degrees of freedom. In either q1D or q2D, a study of the universal dynamics and thermodynamics would be an interesting comparison to true s-wave systems. 

Using a multi-channel framework, new universal contact relations should be found to clarify, for instance, the correspondence to the closed-channel fraction. Qualitatively, a q1D (or q2D) polarized gas with emergent s-wave interactions is expected to be similar to a spin mixture with true s-wave interactions. However for systems such as metastable helium \cite{Jeltes2007,Truscott2023} or dipolar gases with strong spin-orbit relaxation \cite{Chomaz2023}, spin polarization is necessary for collisional stability. For fermionic isotopes in these systems, emergent s-wave interactions in the quasi-low-dimensional regime could provide a new pathway to low-energy universality. 

The paradigm introduced here can be extended to higher partial waves. For example, it could be used to study bosons with d-wave interactions \cite{Cui2017,Yao2019,Fouche2019}. 
The d-wave generalization of Eq.~\eqref{eq:int_q1d} would have $m=0$, $|m|=1$, and $|m|=2$ components. 
The $|m|=1$ matrix element would be proportional to $kk'/a_\perp^4$, and would enable odd-wave scattering, i.e., emergent q1D p-wave interactions between identical bosons, which is forbidden by exchange symmetry in true 1D systems. 
The even-$m$ matrix elements would be proportional to $1/a_\perp^6$, which is in contrast to free space, where low-energy d-wave scattering is suppressed at low $k$. Analogous to the emergent q1D s-wave observed here, strong confinement enhances coupling by the high momentum of the confined wave function. 

In summary, our study provides a comprehensive understanding of the behavior of quasi-one-dimensional fermions interacting with short-range p-wave interactions. Our observations and analyses reveal that multi-orbital scattering in strongly confined systems opens the door to the exploration of alternate exchange symmetries in low-energy scattering channels. The energy-dependence of higher-partial-wave scattering is dominated by the inter-orbital gap, which offers a new approach to suppressing finite-range corrections and studying universal many-body phenomena. Our work indicates the potential of cold-atom experiments with fully activated orbital, spin, and mass degrees of freedom, whose interplay gives rise to a wealth of phenomena in electronic materials. 

\begin{acknowledgments} 
The authors would like to thank S.\ Smale for his early contributions to the experiment, Xiwen Guan and Xiangguo Yin for discussion of Bethe ansatz calculations, and D.\ Cl\'ement for discussions. 
This research is supported by NSERC, AFOSR FA9550-19-1-7044 and FA9550-19-1-0365, ARO W911NF-15-1-0603, HK GRF 17304820, 17304719 and CRF C6009-20G and C7012-21G. 
\end{acknowledgments}

\newpage

\bibliography{pwave}

\begin{appendix} 

\section{Q1D limit of p-wave scattering \label{app:theory}}

To incorporate the correct momentum dependence of p-wave scattering, a model inter-atomic pseudo-potential $U(y,\rho)$ can be written in the following way. Given two arbitrary two-body states $\phi$ and $\chi$, the matrix element of $U$ is 
\be \label{eq:Umatrix}
\langle \phi | U | \chi\rangle = g\boldsymbol{\nabla}\phi^*({\bm 0})\cdot\boldsymbol{\nabla}\chi({\bm 0})\, ,\ee 
evaluated at zero separation between the two fermions, where $g$ is a coupling constant \cite{Kanjilal:2004jv,Pricoupenko:2006,Idziaszek:2009,Gao:2015hz,venu2023unitary}. We note that the potential, as written, is isotropic. In the q1D regime, the relevant two-particle states are denoted as $\ket{\nrad,m,k}$, and the matrix elements of $U$ are 
\begin{align}
\langle \nrad',m',k' | U & | \nrad,m,k \rangle = \frac{g}{\pi a_{\perp}^2} k k' \delta_{m,0} \delta_{m,m'} \nonumber \\
&+ \frac{2g}{\pi a_{\perp}^4}\sqrt{\nrad+1}\sqrt{\nrad'+1} \delta_{|m|,1}\delta_{m,m'}\,.
\label{eq:int_q1d}
\end{align}
where $\hbar$ and atomic mass factors have been suppressed. Since the relative wave function vanishes as $\Psi_{\nrad,m} \propto \rho^{|m|}$, this pseudopoential only has non-zero matrix elements for $|m|\leq 1$. The structure of Eq.~\eqref{eq:int_q1d} is as follows. The first term on the r.h.s.\ corresponds to $m=0$, where the gradient operator acts along the $y$ direction (parallel to $k$ and $k'$). This produces the correct form factor for odd-wave scattering in the q1D direction, $kk'$. As a result, it inherits the low-momentum behavior of the 3D p-wave scattering and vanishes at low energy. On the other hand, the second term corresponds to the gradient operator acting along the transverse directions, and as such, requires antisymmetric radial wave functions ($m=\pm 1$) to give rise to a non-vanishing matrix element. This term is proportional to a constant at low $k$ and $k'$, which is a hallmark of s-wave scattering.

The matrix elements in Eq.~(\ref{eq:int_q1d}) carry forward to a similar structure for the low-energy $T$-matrix in the q1D geometry which we take to have the following form
\begin{align}
\langle & \nrad',  m', k'| T_\mathrm{q1D} |\nrad,m,k \rangle = \frac{k k'}{\pi a_{\perp}^2}  T_\mathrm{q1D}^{(0)}(E_\mathrm{rel})\delta_{m,0} \delta_{m,m'} \nonumber \\
&+ \frac{2}{\pi a_{\perp}^4}\sqrt{\nrad+1}\sqrt{\nrad'+1}\; T_\mathrm{q1D}^{(m)}(E_\mathrm{rel}) \delta_{|m|,1}\delta_{m,m'}\,,
\end{align}
where $T_\mathrm{q1D}^{(m)}$ is the $T$-matrix in the q1D geometry with relative angular momentum $m$. To evaluate $T_\mathrm{q1D}^{(m)}$, we make use of its relation to the free-space $T$-matrix, $T_\mathrm{3D}$ \cite{Moore:2004}:
\begin{align}\label{Tq1DT3D}
&T_\mathrm{q1D}^{(m)}(E_\mathrm{rel}) = T_\mathrm{3D}^{(m)}(E_\mathrm{rel}) \\
&+ T_\mathrm{3D}^{(m)}(E_\mathrm{rel})\left[G_\mathrm{q1D}^{(m)}(E_\mathrm{rel})-G_\mathrm{3D}^{(m)}(E_\mathrm{rel})\right] T_\mathrm{q1D}^{(m)}(E_\mathrm{rel})\,,\nonumber
\end{align}
where $G_\mathrm{q1D}^{(m)}$ and $G_\mathrm{3D}^{(m)}$ are the propagators in the q1D geometry and 3D free space, respectively, with definite angular momentum $m$, and $T_\mathrm{3D}=(2\pi/\mu) f_{\mathrm{3D}}$.

Equation (\ref{Tq1DT3D}) can be evaluated analytically for both the odd-wave ($m=0$) and even-wave ($m=\pm1$) sectors with respective solutions given by
\begin{widetext}
\begin{align} \label{eq:T_0}
T_{o}^{-1} &\equiv\left(\frac{1}{\pi a_{\perp}^2}T_\mathrm{q1D}^{(0)}(E_\mathrm{rel})\right)^{-1} \!\! = 
\frac{2\mu }{2} \left[\frac{a_{\perp}^2}{6} \left(\frac{1}{V_\mathrm{3D}} + \frac{2\mu E_\mathrm{rel}}{R_\mathrm{3D}}\right) - \frac{2}{a_{\perp}} \zeta(-\frac{1}{2},-\mathcal{E}_0 - i \varepsilon)\right] \quad\mbox{and} \\
\label{eq:T_1}
T_{e}^{-1} &\equiv\left(\frac{2}{\pi a_{\perp}^4}T_\mathrm{q1D}^{(\pm 1)}(E_\mathrm{rel})\right)^{-1} \!\! = 
\frac{2\mu }{2} \left[\frac{a_{\perp}^4}{12} \left(\frac{1}{V_\mathrm{3D}} + \frac{2\mu E_\mathrm{rel}}{R_\mathrm{3D}}\right) + \frac{a_{\perp}}{2} \zeta(-\frac{1}{2},-\mathcal{E}_1 - i \varepsilon) + \frac{a_{\perp}}{2} \frac{E_\mathrm{rel}}{2\omega_{\perp}}\zeta(\frac{1}{2},-\mathcal{E}_1 - i \varepsilon)\right]\,. 
\end{align}
\end{widetext}
Here $\zeta(s,x)$ is the Hurwitz zeta function, $\varepsilon$ is an infinitesimally small positive quantity, and $\mathcal{E}_m = \left[E_\mathrm{rel}-(|m|+1)\omega_{\perp}\right]/ 2\omega_{\perp} = (ka_{\perp}/2)^2$. We have also defined the appropriate renormalized q1D scattering $T$-matrices $T_{o}$ (odd wave) and $T_{e}$ (even wave) that have a direct correspondence with scattering $T$-matrices in true 1D systems.

The resonance condition for q1D scattering is $a_o^{-1} = 0$ for the odd-wave case or $a_e = 0$ the for even-wave case. In the low-energy regime $\mathcal{E}_m \ll 1$, the bound-state energies for the odd-wave and even-wave sectors are given by
\be \label{eq:EBo}
E_{B,o} = \omega_{\perp} - \frac{\kappa_o^2}{2\mu}\quad \mbox{with}\quad \kappa_o = \frac{1}{a_o}- \kappa_o^2 r_o \ee
and
\be \label{eq:EBe}
E_{B,e} = 2\omega_{\perp} - \frac{\kappa_e^2}{2\mu}\quad \mbox{with}\quad \frac{1}{\kappa_e} = a_e + \kappa_e^2 r_e\,, \ee
where $\kappa_o>0$ and $\kappa_e>0$ are the odd-wave and even-wave bound-state wave numbers, respectively. The binding energies are consistent with true 1D calculations in the shallow-dimer limit ($\kappa a_\perp \ll 1$): for repulsive odd waves $a_o > 0$, $\omega_\perp-E_{B,o} \approx 1/(2\mu a_o^2)$, and for attractive even waves $a_e > 0$, $2\omega_\perp-E_{B,e} \approx 1/(2\mu a_e^2)$.

\section{RF spectral form in the q1D regime \label{app:spectral}}

Here we derive the analytic expressions for the rf transition rate $\Gamma(\omega)$ at high frequency, taking into account the multiple transverse excited bands. From standard arguments \cite{Randeria2010}, the dominant contribution to the rf transfer rate at high frequencies is determined by the short-range part of the many-body wave function, which is essentially the same as the two-body case at zero energy, apart from a normalization constant (which we will introduce as $\gamma_0$ or $\gamma_1$ momentarily) that depends on the many-body physics. Since we have evaluated the two-body $T$-matrix exactly for cylindrically symmetric q1D scattering in Sec.~\ref{ssec:Tmatrix}, we can evaluate Eq.~\eqref{eq:RF_def} in the high-frequency limit~\cite{Braaten:2008tc, Son:2010}, which gives the exact functional dependencies of $\Gamma(\omega)$ apart from overall constants that characterize the many-body states. We find that the rf transfer rate can be written as $\Gamma(\omega)=\sum_{m=-1}^{+1}\Gamma^{(m)}(\omega)$ with
\begin{align}
\Gamma^{(0)}(\omega) & =  \gamma_0\frac{\Omega^2}{8\pi} \frac{(2\omega_{\perp})^{3/2}}{\omega^2} \label{eq:Gamma0} \\
& \times \sum_{\nrad=0}^{\infty} \bigg( \frac{\omega}{2\omega_{\perp}}-\nrad \bigg)^{1/2}\theta \bigg(\frac{\omega}{2\omega_{\perp}}-\nrad\bigg) \nonumber 
\end{align}
and
\begin{align}
\Gamma^{(\pm 1)}(\omega) & = \gamma_1\frac{\Omega^2}{8\pi} \frac{(2\omega_{\perp})^{3/2}}{\omega^2} \label{eq:Gamma1} \\
& \times \sum_{\nrad=0}^{\infty} (\nrad+1)\bigg(\frac{\omega}{2\omega_{\perp}}-\nrad\bigg)^{-1/2}\! \theta \bigg(\frac{\omega}{2\omega_{\perp}}-\nrad\bigg)\,. \nonumber
\end{align}
Equations~\eqref{eq:Gamma0} and \eqref{eq:Gamma1} are valid for large positive detuning $\omega \gg E_F$. For both $\Gamma^{(0)}(\omega)$ and $\Gamma^{(\pm 1)}(\omega)$, the sum over $\nrad$ represents the contributions from all radially excited states. The physical interpretation of this expression is similar to Eqs.~\eqref{eq:Gamma02} and \eqref{eq:Gamma12} given in the main text, but now generalized: the step function $\theta(\omega/2\omega_{\perp}-\nrad)$ indicates the opening of the transverse band for which the rf photon energy exceeds the energy difference to the $\nrad$-th scattering continuum. 

This calculation clarifies the condition in which true-1D scaling in the rf spectrum is observed. Specifically, the 1D odd- and even-wave contact relations apply when $|\omega|<2\omega_\perp$, a regime in which only the continuum states with $\nrad=0$ are relevant. 
In this case,  
\begin{align}
\Gamma^{(0)}(\omega) &=\gamma_0 \Omega^2\frac{2\omega_\perp}{8\pi}\frac{1}{\omega^{3/2}}
\quad\mbox{and} \\
\Gamma^{(\pm 1)}(\omega) &=\gamma_1 \Omega^2\frac{(2\omega_\perp)^{2}}{8\pi}\frac{1}{\omega^{5/2}}\,.
\end{align}
These two relations should be identified with the expressions for $\Gamma(\omega)$ from the exact 1D odd-wave \cite{Cui:2016kd} and even-wave scattering calculations:
\be \label{eq:Contacts}
\Gamma^{(0)}(\omega) =\frac{\Omega^2}{2\omega^{3/2}} C_{o} \quad \mbox{and} \quad
\Gamma^{(\pm 1)}(\omega) =\frac{\Omega^2}{2\omega^{5/2}} C_e\,, 
\ee
where $C_o$ and $C_e$ are the 1D odd-wave and even-wave contacts, respectively. We thus identify $\gamma_0$ and $\gamma_1$ as
\be
\gamma_0 = \frac{4\pi }{2\omega_{\perp} } C_{o} \quad\mbox{and}\quad 
\gamma_1 = \frac{4\pi}{(2\omega_{\perp})^{2} } C_{a}\,.
\ee
Using these relations in Eqs.~\eqref{eq:Gamma0} and \eqref{eq:Gamma1}, we obtain the limiting forms
Eqs.~\eqref{eq:Gamma02} to \eqref{eq:GamEven1D} in the main text, which are relevant to q1D physics. However, in the limit $\omega\to \infty$, increasingly shorter length scales are probed as the continuum final states of higher transverse bands are made available, eventually approaching the 3D limit; this behavior is also present in Eqs.~\eqref{eq:Gamma0},\eqref{eq:Gamma1}. Replacing the sum over transverse radial states with an integral, one reproduces the correct leading order $\omega^{-1/2}$ rf spectral law for a p-wave Fermi gas in 3D. A related discussion of the q1D momentum distribution is presented in Ref.~\cite{Mingyuan:2021}.

\section{Final-state effects in even-wave spectroscopy \label{app:finalstate}}

The ideal scenario of rf spin-flip spectroscopy is that an atom is transferred from an interacting state to a non-interacting state. This is a good approximation when interactions in the initial spin state are near-resonant, and the final states at the same magnetic field are far from resonance. In the situation probed in our experiments, atoms are spin-flipped from $\ket{b}$ to $\ket{c}$, so that $bb$ p-wave interactions should be large compared to $bc$ p-wave interactions. Coupled-channels calculations indicate that the $z$-dimer $bc$ p-wave Feshbach resonance is located at $B_0 = 217.12$\,G, with a scattering volume that can be parameterized as $V_{bc}(B)/V_\mathrm{bg} = 1-\Delta/(B-B_0)$, with $V_\mathrm{bg}=(-107.95 a_0)^3$ and $\Delta = -8.98$\,G where $a_0$ is the Bohr radius. This yields a zero-crossing near $208$\,G, such that $V_{bc}$ is below its background value in the range of fields across which we measure q1D contacts, typically $200$\,G -- $205$\,G, and can hence be considered as weakly interacting in this regime.

For even-wave spectroscopy, adding final states effects to Eq.~\eqref{eq:GamEven1D}, the relation between $\Gamma$ and $C_e$ is
\be\label{eq:finalstateFULL}
\lim_{\omega\rightarrow\infty} \Gamma(\omega) = \frac{\Omega^2}{2 \sqrt{\hbar m}}\frac{C_e}{\omega^{5/2}} \frac{\left(a_{e} - a_{e}' - m \omega(r_{e}-r_e')/\hbar\right)^2}{(m\omega r_e'/\hbar - a_e')^2 + (m\omega/\hbar)^{-1}}\,,\ee
where $a_e'$ and $r_e'$ are the 1D scattering parameters for the $bc$ channel. 
Equation~\eqref{eq:finalstateFULL} predicts a deviation of $\Gamma(\omega)$ from its asymptotic form, however within the  range of frequency and scattering parameters relevant to the experiment, it does not explain the deviation of the even-wave rf spectrum from the $-5/2$ scaling shown in Fig.~\ref{fig:evenspectrum}.

In our regime of interest, Eq.~\eqref{eq:finalstateFULL} can be approximated as 
\be\label{eq:finalstate}
\lim_{\omega\rightarrow\infty} \Gamma(\omega) = \frac{\Omega^2}{2 \sqrt{\hbar m}}\frac{C_e}{\omega^{5/2}} \frac{\left(a_{e} - a_{e}'\right)^2}{a_e'^2}\,.\ee
This allows us to bound the final-state effects to corrections between $0.90$ and $1.05$ using $V_{bc}$ and a typical value of $R_{\textrm{3D}}$ in the field and depth range used in Fig.~\ref{fig:evencontact}. 
Since this is less than the scatter of the experimental data, we can safely neglect final-state effects. 

\section{High-temperature expansion for the odd- and even-wave contacts \label{app:hte}}

In the high-temperature regime of small fugacity, $z \ll 1$, analytic forms of the many-body contact can be found from a virial expansion \cite{Yu2012,Hofmann2012,Zhang2017,Hu2018,Maki:2021}. Although we explore systems with $T$ comparable to $T_F$, the high-temperature approximation provides a qualitative comparison between theory and experiment with no adjustable parameters. Here we consider a homogeneous 1D system with length $L$, uniform density $n$, and total number $N = L n$. Comparison to experimental data uses a correction for inhomogeneity as described in App.~\ref{app:inhomogeneity}. 

\subsection{Odd-wave contact}

The high-temperature calculation of the odd-wave contact comes from a two-channel model of interacting atoms and closed-channel molecules \cite{Maki:2021}. For weak attractive interactions $|a_o| / \Lambda_T \ll 1$ and $a_o < 0$, 
\be \label{Clweak}
  C_o = \frac{2 \pi \hbar^2 Nn a_o^2}{m \Lambda_T^2}
  \left[1 - 6\pi \frac{a_o r_o}{\Lambda_T^2}\right]\,,
\ee
where $\Lambda_T=(2\pi\hbar^2\beta/m)^{1/2}$ is the single-particle thermal de Broglie wavelength. 
As assumed in App.~\ref{app:inhomogeneity}, the contact is $N n$ times a function of $T$, $a_o$, and $r_o$. A convenient form uses $k_F$ and $E_F$ of the central tube as reference values:
\be
  \frac{C_o}{N} = \frac{\hbar^2 k_F^3 \widetilde T a_o^2}{2 \pi m} \left(1 + \frac32 k_F^2 r_o a_o \widetilde T \right)\,,
\ee
where we have used $\Lambda_T^{-2} =\widetilde T k_F^2/4\pi$ and $k_F = \pi n$.

\subsection{Even-wave contact}

The calculation technique for the even-wave contact follows the high-temperature treatment for odd waves \cite{Maki:2021}, i.e., using a two-channel model of spin-half fermions coupled to closed-channel molecules to evaluate the contacts for arbitrary interaction strengths. To model the anisotropic nature of the interactions and the trapping potential, we assume that the $x$ and $z$ resonances are independent 1D systems. The expression for the contact $C_e^{x,z}$ has a contribution from the scattering continuum ($C_e\vert_\mathrm{at}$) and a contribution from the two-body bound state ($C_e\vert_\mathrm{mol}$). Omitting the direction label $(x,z)$, these two terms are
\be \label{eq:caintatom} C_e\big\vert_\mathrm{at} = 
N_1 n p_0 
\int^\infty_0 \!\!du \frac{2\pi^{-1/2}u^{1/2} e^{-u} }{\frac{m u}{\hbar^2} \big(a_e - \frac{m u}{\hbar^2 \beta} r_e \big)^2+\beta}\,, \ee
where the integration variable is dimensionless, $p_0 \equiv N_0/(N_0+N_1)$ is the population fraction of ground-band atoms, and
\be \label{eq:caintmol} 
C_e\big\vert_\mathrm{mol} \approx 
N_1 n p_0 \frac{\hbar^2}{m} \frac{\sqrt{2}\Lambda_T \, \theta(a_e)}{r_e + |a_e|^3/2} e^{\beta \hbar^2/(ma_e^2)} \,,\ee
where $\theta()$ is the Heaviside step function. Note that in Eq.~(\ref{eq:caintmol}), we have used the shallow-dimer limit instead of the full Eq.~\eqref{eq:EBe}, which is a good approximation in the $k_F a_e > 2$ range for which the high-temperature $C_e$ is considered in Fig.~\ref{fig:evencontact}. Just like the odd-wave case, the even-wave contact is a function of $N n$, but now multiplied by the fraction in each band. 

Scaling by $k_F$ and $E_F$ of the central tube, the sum of these contributions is
\begin{align} \label{eq:Caint}
\frac{C_e}{N_1} = & \frac{2p_0 \hbar^2 k_F^3}{\pi^{3/2}m} \bigg\{ \frac{\sqrt{2}\pi e^{2 /(k_F^2 a_e^2 \widetilde{T})} }{k_F^3 r_e + |k_F a_e|^3/2} \widetilde{T}^{-1/2} \nonumber \\
& + \int^\infty_0 \!\!du \frac{ u^{1/2}e^{-u} }{u (k_F a_e - u k_F^3 r_e \widetilde{T}/2)^2 + 2 \widetilde{T}^{-1}} \bigg\}\,, \end{align}
which is plotted in Fig.~\ref{fig:evencontact}. In the weakly interacting limit the even-wave contacts are given by
\be
\frac{C_e}{N_1} \approx \frac{2 p_0 \hbar^2 k_F}{\pi m} \, a_e^{-2}
\ee

In the resonantly interacting limit, the high temperature theory breaks down, as the binding energy $ \propto 1/a_e^2$ quickly exceeds the temperature, invalidating the high-temperature assumption. 

\begin{figure*}[t!]
\centering
\includegraphics[width=2\columnwidth]{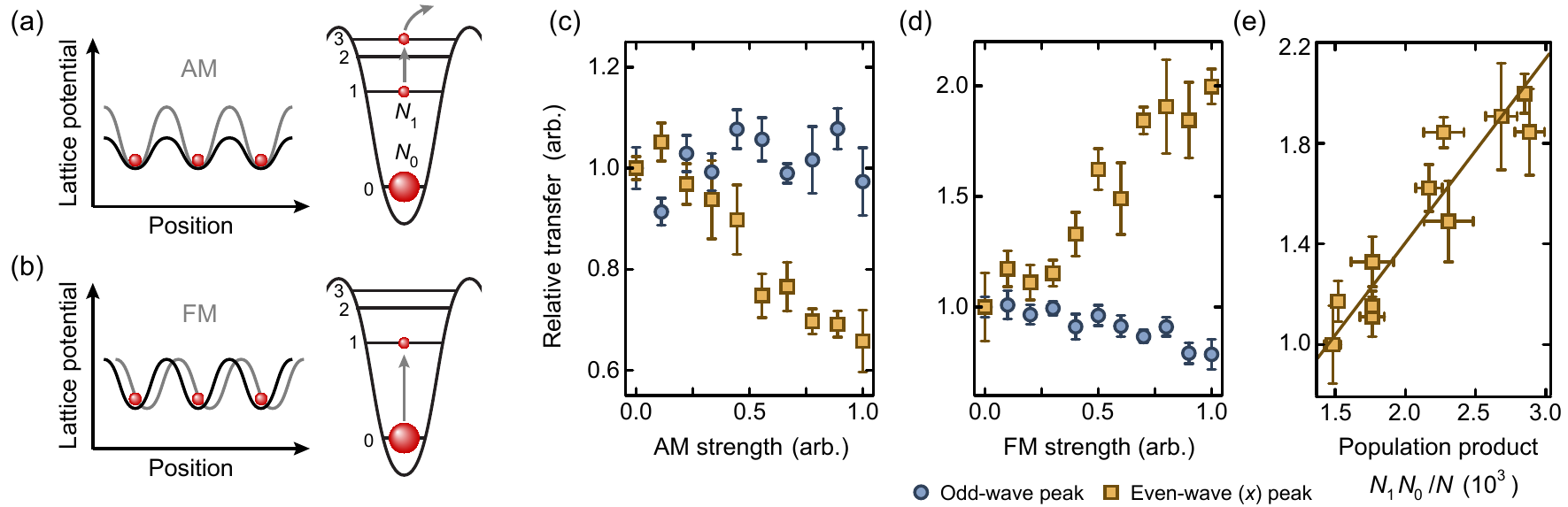}
\caption{{\em Correlation of transfer with modulation type and strength}. (a) Amplitude modulation of the lattice potential causes a single-particle first-to-third transition that removes atoms from the first excited band. (b) Frequency modulation causes a single-particle zero-to-first transition to populate the excited band while also depleting the ground band. (c) Number of transferred atoms to probe state from $\widetilde{\omega}/2\pi = 50$\,kHz for $40\,E_R$ confinement with varying AM strength. No significant change in transfer at the odd-wave peak ($199.25$\,G). Less transfer is noticed at the even-wave ($x$) peak ($199.75$\,G). Transfer is plotted relative to the transfer with no modulation. (d) Transferred atoms for various frequency modulation. A slight decrease in transfer for the odd-wave peak is seen as atoms are transferred out of the ground band. Transfer is increased for the even-wave peak. (e) Transfer at even-wave peak is plotted versus the population product excited- and ground-state populations (see text) divided by $N_\mathrm{tot}$, and compared to a linear fit.} 
\label{fig:9} 
\end{figure*}

\section{RF spectroscopy with variable band populations \label{app:bandpop}}

The odd- and even-wave contacts are extensive quantities that are proportional to population in the relevant bands at high temperature. We are able to manipulate band populations with two types of lattice modulations, amplitude modulation (AM) and frequency modulation (FM). In this Appendix, we show their effects on the measured contacts. 

Figure~\ref{fig:9}(a) illustrates AM and its primary effect on band populations. Anharmonicity of the lattice allows for the tuning of the frequency of modulation to be such that only one transition is resonant: AM frequency resonant first-to-third excited states transfer is off-resonant for excitation from the ground band to the second excited band. Since atoms in the third band leave the trap, this process distills the ground band. 

Phase modulation of the optical standing wave is implemented experimentally by FM of the light, which creates a time-dependent displacement of local potential minima where atoms are trapped, due to the optical propagation to the retro-reflecting mirror and back. Figure~\ref{fig:9}(b) illustrates this local position displacement of the lattice potential. Unlike AM, phase modulation provides the necessary parity-breaking time-dependent gradient to transfer atoms from the ground band to the first excited band. 

The effect of AM on the odd- and even-wave contacts are shown in Fig.~\ref{fig:9}(c). Here circles and squares report $\Gamma$ at magnetic fields near the odd-wave and even-wave peaks, respectively, which is proportional to the (extensive) contact. We see that increasing AM strength has no noticeable trend in $\Gamma$ at the odd-wave peak, which is consistent with odd-wave correlations being due to scattering between ground-band atoms. At the even-wave peak, decrease of $\Gamma$ with AM is also consistent with even-wave correlations being proportional to the exited-band population. 

Similarly, the effect of FM is shown in Fig.~\ref{fig:9}(d). Since FM modulates atoms from the ground band to the excited band, one might expect an equal change in the strength of correlations for odd and even-waves. Instead, we observe that increasing modulation strength only slightly decreases transfer at the odd-wave peak and causes comparatively large increases in transfer for the even-wave peak. From this we remark that the relatively equal weights for the peaks seen in Fig.~\ref{fig:results}(b) is coincidental for the $\sim 15\%$ of excited-band atoms populated by our multi-band loading procedure. 

Figure~\ref{fig:9}(e) demonstrates how transfer at the even-wave peak is correlated with the product of ground- and excited-state atom populations, counted in band-mapped time-of-flight imaging. A linear fit of the measurements has a zero crossing within uncertainty of zero and a reduced $\chi^2$ of 1.07. This measurement provides us with the confidence to attribute the transfer seen at the even-wave peaks as a measure of correlations between ground and excited band atoms. 

\section{Inhomogeneous averaging \label{app:inhomogeneity}}

The theoretical description of atom-atom correlations throughout this manuscript assume a single q1D system with a homogeneous atomic density. Our experimental measurements average across two types of inhomogeneity: the density profile within a single tube, and tube-to-tube variation in $N$. 
In the high-temperature limit, $C_\mathrm{hom}/Nn_\textrm{1D}$ for a uniform-density gas is solely a function of temperature, scattering parameters, and confinement strength (see Sec.~\ref{app:hte}), i.e., variables that are uniform across our ensemble. We can thus account for inhomogeneities by separating out the dependence of contact on atom number and density, as
\be \langle C \rangle_\mathrm{obs} = \langle N n_\textrm{1D} \rangle \frac{C_\mathrm{hom}}{N n_\textrm{1D}} \ee
where $\langle N n \rangle$ is an average across the ensemble.

The inhomogeneity within each tube is treated with a local density approximation: that each differential length $dy$ contributes $n_\textrm{1D}^2(y) dy$ to $N n_\textrm{1D}$. 
In the high-temperature limit, the density distribution is given by a Maxwell-Boltzmann distribution, so 
\begin{equation} \label{eq:Nn_tube}
   \langle N n_\textrm{1D} \rangle_\mathrm{tube} = \int n_{\textrm{1D}}^2(y) dy = (4 \pi)^{-1/2} {N^2}/{R_T} 
 \end{equation}
where $R_T = (k_B T /m \omega_y^2)^{1/2}$ is the thermal length of the cloud. By comparison, the peak 1D density is $(2 \pi)^{-1/2} N / R_T$, such that $\langle N n_\textrm{1D} \rangle_\mathrm{tube} = N n_\mathrm{pk}/\sqrt{2}$.

We model the distribution of q1D ensembles by assuming the transfer between a 3D Fermi gas and an array of tubes is isentropic. This approach neglects the fraction of atoms that appear in higher bands, since edges of the Brillouin zone cannot be adiabatic. 
This transfer occurs during the first $t_\mathrm{ramp}$ phase of Fig.~\ref{fig:methods}(b)(i), up to a modest depth ($V_L < 10 E_R$). 
The density of states of the low-depth array is a convolution of a harmonic-lattice density of states $\rho(\epsilon) = 2\sqrt{E_R}/\pi\hbar \omega_\perp \sqrt{\epsilon}$ \cite{rey:2005,blakie:2007} in the $x$ and $z$ directions with a 1D harmonic oscillator $\rho(\epsilon) = 1/\hbar \omega_y$ along the tube where $\epsilon$ is energy. 
We find the chemical potential and temperature of the transformed tube array by imposing entropy and number conservation between the pre-loaded gas to the array of tubes. 
After this loading phase, energy and particle transport between the tubes are suppressed as the lattice depths increase to their final values. 
Including a $-20\%$ corrective factor for atom loss observed in round-trip loading calibrations, 
we find the number and temperature of the central tube to be $N^* \approx 30$ and $\widetilde T \approx 0.8$ for typical conditions, where $\widetilde T = k_B T/E_F$; here and below, we will use $N^*$ to define a reference value of the 1D Fermi energy $E_F$. 
The atom-number distribution of this adiabatic loading model is compared to a diabatic model in which the initial 3D density distribution is partitioned into an square array of parallelepipeds. 
In both models, the distribution of $N$ is found to be roughly uniform: i.e., as many tubes have near-peak atom number as have only one atom. 

We define a distribution function $w_N$ of atom number that satisfies $\sum_N w_N = M$, where $M$ is the total number of tubes, and $\sum_N w_N N = N_{\mathrm{tot}}$. 
Without a dispersion in atom number, each of the $M$ tubes would have $N^*$ atoms, so that $N_\mathrm{tot}=N^*M$, and the distribution function would be $w_N=N_\mathrm{tot}\delta_{N,N^*}/N^*$.
Instead, in the distribution found by these loading models, $w^{\mathrm{exp}}_N$ is approximately constant between $N=1$ and $N=N^*$, and zero otherwise. This gives
\begin{equation}
  N_{\mathrm{tot}} = w^{\mathrm{exp}}_N \sum_{N=1}^{N^*} N = \frac{1}{2}(N^* + 1)N^* w^{\mathrm{exp}}_N\,.
\end{equation}
We can rearrange this expression to find $w^{\mathrm{exp}}_N = 2N_{\mathrm{tot}}/[(N^*+1)N^*]$ for $1 \leq N \leq N^*$. Since the contact is proportional to $N^2$ in the high-temperature limit [see Eq.~\eqref{eq:Nn_tube}], the ratio of the contacts in these two cases is given by 
\begin{equation}
  \frac{\langle N^2 \rangle^{\mathrm{exp}}}{(N^*)^2} = \frac{1}{N_\mathrm{tot} N^*} \sum_{N=1}^{N^*} w^{\mathrm{exp}}_N N^2 =
  \frac{1+2N^*}{3N^*} \,.
\end{equation}
which approaches $2/3$ for large $N^*$.

Combining the intra-tube and inter-tube inhomogeneity effects yields a correction in the estimated value of the contact 
\begin{equation} \label{eq:inhomogeneity}
  \frac{\langle C_\mathrm{tube} \rangle^{\mathrm{exp}}}{ C_\mathrm{hom}} = \frac{2}{3}\sqrt{\frac{\pi}{8}}\widetilde{T}^{-1/2}\,.
\end{equation}
The theoretical curves shown in Figs.~\ref{fig:oddcontact} and \ref{fig:evencontact} use the uniform-gas prediction multiplied by this correction, which is $\sim 0.4$ for typical loading parameters. 

\end{appendix}

\end{document}